\documentclass[final,twocolumn,5p,times,10pt,authoryear]{elsarticle}
\usepackage{lineno}
\modulolinenumbers[5]

\journal{Automatica}

\usepackage[american]{babel}
\usepackage[utf8]{inputenc}
\usepackage[T1]{fontenc}
\usepackage{graphicx}
\usepackage{amsmath}
\usepackage{amssymb}
\usepackage{amsthm}
\usepackage{amsfonts}
\usepackage{bm}
\usepackage{hhline}
\usepackage{csquotes}
\usepackage{color}
\usepackage{xcolor}
\usepackage{import}
\usepackage{algorithm}
\usepackage{algpseudocode}
\usepackage{dsfont}
\usepackage{multirow}
\usepackage{soul}
\setstcolor{magenta}
\newcommand{\GR}[1]{\textcolor{magenta}{#1}}
\renewcommand{\GR}[1]{}

\makeatletter
\usepackage[colorlinks=true,linkcolor=red]{hyperref}
\makeatother

\newtheorem{theorem}{\bf{Theorem}}
\newtheorem{lemma}{\bf{Lemma}}
\newtheorem{proposition}{\bf{Proposition}}
\newtheorem{definition}{\bf{Definition}}
\newtheorem{remark}{\bf{Remark}}

\newcommand{\mbf}[1]{\ensuremath{{\mathbf{#1}}}}

\newcommand{\ten}[1]{\ensuremath{{\cdot}10^{#1}}}

\newcommand{\eye}[1]{\ensuremath{\mbf{I}_{#1}}}
\newcommand{\eyenoarg}{\ensuremath{\mbf{I}}}
\newcommand{\zeros}[2]{\ensuremath{\bm{0}_{#1\times#2}}}

\newcommand{\real}[1]{\ensuremath{\text{Re}(#1)}}

\newcommand{\realset}{\ensuremath{\mathbb{R}}}
\newcommand{\realsetmat}[2]{\ensuremath{\mathbb{R}^{#1\times#2}}}

\newcommand{\naturalset}{\ensuremath{\mathbb{N}}}

\newcommand{\modelset}{\ensuremath{\mathcal{M}}}

\newcommand{\ninf}[1]{\ensuremath{\|{#1}\|_\infty}}

\definecolor{orange}{RGB}{255,69,0}
\newcommand{\textred}[1]{\textcolor{red}{#1}}

\renewcommand{\textred}[1]{#1}

\newcommand{\seq}[1]{\ensuremath{{\protect\overset{\to}{#1}}}} 

\newcommand{\zerospace}{\ensuremath{{\,\!}}}

\newcommand{\zspace}{\ensuremath{\zerospace}}
\newcommand{\inv}[1]{\ensuremath{{#1}\zerospace^{-1}}}

\newcommand{\zseq}[1]{\ensuremath{{\protect\overset{\to}{#1}}\zspace}} %

\newcommand{\noarg}{\ensuremath{\_\,}}
\newcommand{\emptyarg}{\noarg}

\newcommand{\cmrk}{\ensuremath{\checkmark}}

\newcommand{\mixedball}[3]{\ensuremath{B_{\realset,\infty}({#1},{#2},{#3})}}

\newcommand{\zon}{\ensuremath{_\text{Z}}}
\newcommand{\czon}{\ensuremath{_\text{CZ}}}

\newcommand{\lzon}{\ensuremath{_\text{LZ}}}

\newcommand{\bdiag}[1]{\ensuremath{\text{bdiag}({#1})}}
\newcommand{\verti}[1]{\ensuremath{\text{verti}({#1})}}

\newcommand{\bibfolder}{Bibliography}

\biboptions{authoryear}

\allowdisplaybreaks 

\begin{document}

\begin{frontmatter}

\title{Line zonotopes: A tool for state estimation and fault diagnosis of unbounded and descriptor  systems\tnoteref{mytitlenote}}

\tnotetext[mytitlenote]{This work was partially supported by the Brazilian agencies CNPq, under grants 465755/2014-3 (INCT project), 317058/2023-1 and 422143/2023-5; FAPESP, under grants 2014/50851-0, and 2022/05052-8; CAPES through the Academic Excellence Program (PROEX); FAPEMIG; and by the CHIPS Joint Undertaking and the European Union through the EcoMobility project (grant 101112306). 
}

\author[myfirstaddress]{Brenner S. Rego\corref{mycorrespondingauthor}}
\cortext[mycorrespondingauthor]{Corresponding author, \textit{brennersr7@usp.br} }

\author[mysecondaddress]{Davide M. Raimondo}

\author[mythirdaddress]{Guilherme V. Raffo}

\address[myfirstaddress]{Department of Electrical and Computer Engineering, University of São Paulo, São Carlos, SP 13566-590, Brazil}
\address[mysecondaddress]{Department of Engineering and Architecture, University of Trieste, 34127, Trieste, Italy}
\address[mythirdaddress]{Department of Electronics Engineering, Federal University of Minas Gerais, Belo Horizonte, MG 31270-901, Brazil}

\begin{abstract}
This paper proposes new methods for set-based state estimation and active fault diagnosis (AFD) of linear descriptor systems (LDS). Unlike intervals, ellipsoids, and zonotopes, constrained zonotopes (CZs) can directly incorporate linear static constraints on state variables -- typical of descriptor systems -- into their mathematical representation, leading to less conservative enclosures.  However, for LDS that are unstable or not fully observable, a bounded representation cannot ensure a valid enclosure of the states over time. To address this limitation, we introduce line zonotopes, a new representation for unbounded sets that retains key properties of CZs, including polynomial time complexity reduction methods, while enabling the description of strips, hyperplanes, and the entire $n$-dimensional Euclidean space. This extension not only generalizes the use of CZs to unbounded settings but can also enhance set-based estimation and AFD in both stable and unstable scenarios. 
Additionally, we extend the AFD method for LDS from \cite{Rego2020b} to operate over reachable tubes rather than solely on the reachable set at the final time of the considered horizon. This reduces conservatism in input separation and enables more accurate fault diagnosis based on the entire output sequence. The advantages of the proposed methods over existing CZ-based approaches are demonstrated through numerical examples.
\end{abstract}



\end{frontmatter}


\section{Introduction}

Descriptor systems (also known as singular or implicit systems) exhibit dynamic and static behaviors described through differential and algebraic equations \citep{Puig2018}. These characteristics are present in many physical systems of practical interest, such as socioeconomic systems, chemical processes \citep{Kumar1995}, battery packs, water distribution networks \citep{Vrachimis2018c}, and robotic systems with holonomic and nonholonomic constraints \citep{JAN11,Yang2019}. However, their unique characteristics make descriptor systems particularly challenging for control, state estimation, and fault diagnosis.

Set-based state estimation is a well-established topic in the literature, first addressed in the 1960s \citep{Schweppe1968}, and is particularly important for systems with bounded uncertainties. This holds true for linear descriptor systems (LDS) as well. Luenberger-type observers have been used in \cite{Merhy2019} for state estimation of LDS, with sets described by ellipsoids. Despite providing stable bounds, ellipsoids are conservative since the complexity of the sets is fixed, leading to a significant wrapping effect. Due to their properties and arbitrary complexity, zonotopes have been used in more recent methods for state estimation of LDS \citep{Puig2018,Wang2018}. However, zonotopes are not able to incorporate linear equality constraints, which are related to the static behavior present in descriptor systems. To address this limitation, a state estimation method for LDS based on \emph{constrained zonotopes} (CZs) was proposed in \cite{Rego2020b}. Constrained zonotopes can capture the equality constraints inherent of LDS, while maintaining many of the properties of zonotopes. Nevertheless, the proposed CZ-based method requires prior knowledge of an admissible state set enclosing all system states throughout the entire experiment, which is an unrealistic assumption in the case of unstable  LDS.
 
To address these limitations, this manuscript introduces \emph{line zonotopes} (LZs), an extension of CZs designed to describe unbounded sets such as strips, hyperplanes, and the entire $n$-dimensional Euclidean space. LZs inherit key properties from CZs, such as efficient Minkowski sum, linear mapping, and intersection while maintaining efficient complexity reduction methods. 
This latter represents an important aspect of LZs, which is not provided by other extensions, such as constrained polynomial zonotopes \citep{Kochdumper2023CPZ}, hybrid zonotopes \citep{Bird2023}, and constrained convex generators (CCGs) \citep{Silvestre2022}. Moreover, unlike ellipsotopes \citep{Kousik2023Ellipsotopes}, LZs allow the application of fully developed reduction methods to the general case.  LZs can be used in set-based state estimation eliminating the need for prior knowledge of bounds on the initial state or subsequent trajectories. This allows tackling unstable systems where CZ methods are shown to fail.  Moreover, LZs are advantageous for addressing unobservable LDS, where the absence of measurements prevents providing initial bounds for some states. 
 
Another important topic addressed in this work is set-based fault diagnosis of LDS. The goal is to identify which fault is occurring in the system. Zonotopes have been used in \cite{Wang2019b} to identify additive faults using unknown input observers, and in \cite{Yang2019} to investigate active fault diagnosis (AFD) of LDS through the separation of reachable sets using a designed input sequence. However, as in state estimation, zonotopes cannot capture effectively the static relationships of descriptor systems, requiring a longer time interval for fault diagnosis. An AFD method using CZs has been proposed in \cite{Rego2020b} based on the separation of reachable output sets using mixed-integer quadratic programming. This approach is able to incorporate the linear equality constraints of LDS. Still, the proposed method requires the knowledge of an admissible state set, which is a prohibitive assumption for online applications involving unstable and unobservable systems.

In this work, we also extend the open-loop AFD method \cite{Rego2020b} using LZs, thus allowing the separation of unbounded sets,  but also operating over reachable tubes rather than solely on the reachable set at the final time of the considered horizon. By leveraging the entire output sequence, this method reduces the required input length and norm for fault isolation. Note that both the AFD method in \cite{Rego2020b} and the new method in this paper introduce input-dependent constraints, making complexity reduction nontrivial. This challenge is specifically addressed here.
 
The effectiveness of the proposed methods with respect to CZ approaches is highlighted in numerical examples. In special, the use of LZs enables exactness and lower computational costs in part of the complexity reduction procedures required for state estimation and AFD of LDS, in comparison to CZs. 

This paper is organized as follows. Mathematical preliminaries on set operations and CZs are described in Section \ref{sec:desc_problemformulation}, along with the problem formulation and a motivational example. Section \ref{sec:linezonotopes} introduces the set representation proposed in this paper, called line zonotopes, along with their fundamental properties, complexity reduction methods, and conceptual examples. The new state estimation method based on LZs is described in Section \ref{sec:desc_estimationLZ}. Section \ref{sec:AFDLZ} develops the tube-based AFD method using line zonotopes. Numerical examples are presented in Section \ref{sec:results} to corroborate the effectiveness of the respective new methods, and Section \ref{sec:conclusions} concludes the paper.

\section{Preliminaries and problem formulation} \label{sec:desc_problemformulation}

\emph{Constrained zonotopes}, as defined in \cite{Scott2016}, extend the concept of zonotopes by enabling the representation of asymmetric convex polytopes while preserving many of the computational advantages associated with zonotopes.
\begin{definition} \rm \label{def:pre_czonotopes}
	A set $Z \subset \realset^n$ is a \emph{constrained zonotope} if there exists $(\mbf{G}_z,\mbf{c}_z,\mbf{A}_z,\mbf{b}_z) \in \realsetmat{n}{n_g} \times \realset^n \times \realsetmat{n_c}{n_g} \times \realset^{n_c}$ such that
	\begin{equation} \label{eq:pre_cgrep}
	Z = \left\{ \mbf{c}_z + \mbf{G}_z \bm{\xi} : \| \bm{\xi} \|_\infty \leq 1, \mbf{A}_z \bm{\xi} = \mbf{b}_z \right\}.
	\end{equation}	
\end{definition}

We refer to \eqref{eq:pre_cgrep} as the \emph{constrained generator representation} (CG-rep). Each column of $\mbf{G}_z$ is a line segment, denoted as \emph{generator}, $\mbf{c}_z$ is the \emph{center}, and $\mbf{A}_z \bm{\xi} = \mbf{b}_z$ are the \emph{constraints}. We use the notation $Z = (\mbf{G}_z, \mbf{c}_z,\mbf{A}_z,\mbf{b}_z)\czon$ for CZs, and $Z = (\mbf{G}_z, \mbf{c}_z)\zon$ for zonotopes\footnote{Generator representation, referred to as G-rep.}. Consider $Z, W \subset \realset^{n}$, $Y \subset \realset^{m}$, and $\mbf{R} \in \realset^{m \times n}$. Let $Z \times W  \triangleq \{(\mbf{z},\mbf{w}): \mbf{z} \in Z, ~\mbf{w} \in W\}$ be the Cartesian product, $\mbf{R}Z \triangleq \{ \mbf{R} \mbf{z} : \mbf{z} \in Z\}$ be the linear mapping, $Z \oplus W \triangleq \{ \mbf{z} + \mbf{w} : \mbf{z} \in Z,\, \mbf{w} \in W\}$ be the Minkowski sum, and ${Z \cap_{\mbf{R}} Y} \triangleq \{ \mbf{z} \in Z : \mbf{R} \mbf{z} \in Y\}$ be the generalized intersection. These set operations can be computed exactly and trivially with CZs, while leading to a linear increase in the complexity of the CG-rep.  By defining the constrained unitary hypercube\footnote{We use the notation $B_\infty^{n_g}$ for the $n_g$-dimensional unitary hypercube (i.e., without equality constraints). We drop the superscript $n_g$ for $B_\infty(\mbf{A}_z,\mbf{b}_z)$ since this dimension can be inferred from the number of columns of $\mbf{A}_z$.} $B_\infty(\mbf{A}_z,\mbf{b}_z) \triangleq \{\bm{\xi} \in \realset^{n_g} : \ninf{\bm{\xi}} \leq 1,\,  \mbf{A}_z \bm{\xi} = \mbf{b}_z \}$, a \textred{CZ} $Z$ can be written as 
$Z = \mbf{c} \oplus \mbf{G}_z B_\infty(\mbf{A}_z,\mbf{b}_z)$. Efficient methods for complexity reduction are available to enclose a CZ within another CZ with fewer generators and constraints \citep{Scott2016}.

\subsection{Motivational example} \label{sec:desc_motivational}

In this subsection, we present a motivational example illustrating that, in the general case, the feasible sets of a linear discrete-time descriptor system cannot be described exactly by zonotopes even when $X_0$, $W$, and $V$ are all zonotopes. For the sake of simplicity, consider
\begin{equation}
\begin{aligned} \label{eq:desc_systemmotivational}
\mbf{E} \mbf{x}_k & = \mbf{A} \mbf{x}_{k-1}, \\
\end{aligned}
\end{equation}
with 
\begin{equation*}
\begin{aligned}
& \mbf{E} = \begin{bmatrix} 1 & 0 & 0 \\ 0 & 1 & 0 \\ 0 & 0 & 0 \end{bmatrix}, \; \mbf{A} = \begin{bmatrix} 0.5 & 0 & 0 \\ 0.8 & 0.95 & 0 \\ 0.3 & 0.1 & 0.1 \end{bmatrix}.
\end{aligned}
\end{equation*}

\begin{figure}[!tb]
	\centering{
		\def\svgwidth{0.9\columnwidth}
  {\scriptsize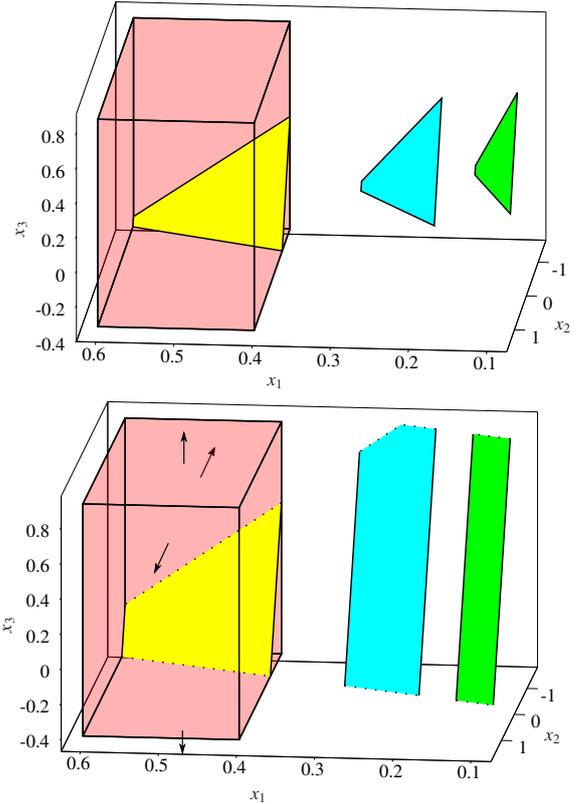}
		\caption{Top: The feasible sets for the system \eqref{eq:desc_systemmotivational}: the initial set $X_0$ (red), and the sets $S_0$ (yellow), $S_1$ (cyan), and $S_2$ (green). Bottom: The feasible sets for the system \eqref{eq:desc_systemmotivational} with $X_0$ (red) unbounded along $x_2$ and $x_3$ (denoted by arrows).}\label{fig:desc_motivational}}
\end{figure}

Let $\mbf{x}_0 \in X_0$, where $X_0$ is a zonotope given by
\begin{equation} \label{eq:desc_motivationalX0}
X_0 = \left(\begin{bmatrix} 0.1 & 0 & 0 \\ 0 & 1.5 & 0 \\ 0 & 0 & 0.6 \end{bmatrix}, \begin{bmatrix} 0.5 \\ 0 \\ 0.25 \end{bmatrix} \right)\zon.
\end{equation}
Since the last row of $\mbf{E}$ is zero, the dynamics \eqref{eq:desc_systemmotivational} further constrains $\mbf{x}_{k-1}$ for $k \geq 1$. In particular, $\mbf{x}_0$ should be consistent with both \eqref{eq:desc_motivationalX0} and the constraints in \eqref{eq:desc_systemmotivational}. Therefore, $\mbf{x}_0 \in S_0$ with $S_0 = \{\mbf{x} \in X_0: \eqref{eq:desc_systemmotivational} \text{ holds for }k = 1\}$. Similarly, we define $S_{k} = \{\mbf{x} \in \realset^n : \mbf{E} \mbf{x}_{k+1} = \mbf{A} \mbf{x},~ \mbf{E} \mbf{x} = \mbf{A} \mbf{x}', \mbf{x}' \in S_{k-1}\}$, with $k \geq 1$. Figure \ref{fig:desc_motivational} (top) shows the zonotope $X_0$ and the convex polytopes $S_k$ for $k = 0,1,2$. These latter are computed exactly using the MPT toolbox \citep{MPT3}. As it can be noticed, the sets $S_k$ are not symmetric, and therefore cannot be described accurately by zonotopes. In addition, due to the scarcity of information in practical systems, the known set $X_0$ may not be bounded in general, resulting in unbounded reachable sets $S_k$ which cannot be described by constrained zonotopes (see Figure \ref{fig:desc_motivational}, bottom).

\subsection{Problem formulation}

Consider a linear discrete-time descriptor system with time $k$, state $\mbf{x}_k \in \realset^{n}$, input $\mbf{u}_{k} \in \realset^{n_u}$, process uncertainty $\mbf{w}_k \in \realset^{n_w}$, measured output $\mbf{y}_k \in \realset^{n_y}$, and measurement uncertainty $\mbf{v}_k \in \realset^{n_v}$. In each discrete-time tuple $(k-1,k)$, $k=1,2,\ldots$, the system evolves according to one of possible $n_m$ known models
\begin{equation}
\begin{aligned} \label{eq:desc_system}
\mbf{E}^{[i]} \mbf{x}_k^{[i]} & = \mbf{A}^{[i]}  \mbf{x}_{k-1}^{[i]} + \mbf{B}^{[i]}  \mbf{u}_{k-1} + \mbf{B}_w^{[i]}  \mbf{w}_{k-1}, \\
\mbf{y}_k^{[i]} & = \mbf{C}^{[i]}  \mbf{x}_k^{[i]} + \mbf{D}^{[i]} \mbf{u}_{k} + \mbf{D}_v^{[i]} \mbf{v}_{k},
\end{aligned}
\end{equation}
with $\mbf{E}^{[i]} \in \realsetmat{n}{n}$, $\mbf{A}^{[i]} \in \realsetmat{n}{n}$, $\mbf{B}^{[i]} \in \realsetmat{n}{n_u}$, $\mbf{B}_w^{[i]} \in \realsetmat{n}{n_w}$, $\mbf{C}^{[i]} \in \realsetmat{n_y}{n}$, $\mbf{D}^{[i]} \in \realsetmat{n_y}{n_u}$, and $\mbf{D}_v^{[i]} \in \realsetmat{n_y}{n_v}$, $i \in \modelset \triangleq \{1,2,\ldots,n_m\}$. We assume that $\mbf{E}^{[i]}$ can be singular. In such a case, one has $n - \text{rank}(\mbf{E})$ purely static constraints. In addition, $\mbf{x}^{[i]}_0 \in X_0$ and $(\mbf{w}_{k},\mbf{v}_{k}) \in W \times V$ for all $k \geq 0$, where $X_0$, $W$ and $V$ are known convex polytopic sets. Moreover, the initial condition $(\mbf{x}^{[i]}_0,\mbf{u}_0,\mbf{w}_0,\mbf{v}_{0})$ is assumed to be feasible, i.e., consistent with the static relations in \eqref{eq:desc_system}. The output $\mbf{y}_0$ is determined based on the initial condition, i.e. $\mbf{y}_0^{[i]} = \mbf{C}^{[i]} \mbf{x}^{[i]}_0 + \mbf{D}^{[i]} \mbf{u}_{0} + \mbf{D}^{[i]}_v \mbf{v}_{0}$. Furthermore, $n_q$ denotes the number of distinct combinations of $(i,j) \in \modelset \times \modelset$, $i \neq j$, and $\mathcal{Q} \triangleq \{1,2,\ldots,n_q\}$. 

Our work has two primary goals. The first, \emph{state estimation}, aims to compute a tight enclosure $\hat{X}_k$ of all possible states that are consistent with \eqref{eq:desc_system} and the given uncertainties for each $k \geq 0$. 
We assume the system follows one of the models $i \in \modelset$ in \eqref{eq:desc_system}, with $i$ known a priori. The second objective, \emph{active fault diagnosis}, is to identify the model that accurately describes the system’s behavior, facilitated by the application of a suitable input $\mbf{u}_k$. The system dynamics are assumed to remain unchanged throughout the diagnosis procedure, meaning that the AFD is performed quickly enough to prevent model switching. Accordingly, a sequence $(\mbf{u}_0, \mbf{u}_1, \dots, \mbf{u}_N)$ of minimal length $N$ is designed to ensure that each possible output sequence $(\mbf{y}_0^{[i]}, \mbf{y}_1^{[i]}, \dots, \mbf{y}_N^{[i]})$ is consistent with only one model $i \in \modelset$. 
Whenever feasible, this problem may have multiple valid input solutions. To address this, we define a cost function and select the optimal input sequence from the feasible set, assuming $\mbf{u}_k \in U$, with $U$ being a known bounded convex polytopic set.

\section{Line zonotopes} \label{sec:linezonotopes}

In this section, we introduce a new set representation called the \emph{line zonotope}, which extends the concept of CZs to encompass a broader class of sets. Line zonotopes retain the advantages of CZs, while also offering the additional capability to describe unbounded sets.

\begin{definition} \rm \label{def:desc_lzonotopes}
	A set $Z \subseteq \realset^n$ is a \emph{line zonotope} if there exists $(\mbf{M}_z,\mbf{G}_z,\mbf{c}_z,\mbf{S}_z,\mbf{A}_z,\mbf{b}_z) \in \realsetmat{n}{n_\delta} \times \realsetmat{n}{n_g} \times \realset^n \times \realsetmat{n_c}{n_\delta} \times \realsetmat{n_c}{n_g} \times \realset^{n_c}$ such that
	\begin{equation} \label{eq:desc_tgrep}
	Z = \{ \mbf{c}_z + \mbf{M}_z \bm{\delta} + \mbf{G}_z \bm{\xi}: \bm{\delta} \in \realset^{n_\delta}, \| \bm{\xi} \|_\infty \leq 1, \mbf{S}_z \bm{\delta} + \mbf{A}_z \bm{\xi} = \mbf{b}_z \}.
	\end{equation}	
\end{definition}

We define \eqref{eq:desc_tgrep} as the \emph{constrained line-and-generator representation} (CLG-rep). In this representation, each column of $\mbf{M}_z$ corresponds to a \emph{line}, each column of $\mbf{G}_z$ represents a \emph{generator}, $\mbf{c}_z$ is the \emph{center}, and $\mbf{S}_z \bm{\delta} + \mbf{A}_z \bm{\xi} = \mbf{b}_z$ define the \emph{constraints}. Unlike generators (which represent line segments), lines are unbounded. Consequently, according to the definition of the CLG-rep in \eqref{eq:desc_tgrep}, if $\mbf{S}_z$ is a matrix of zeros, then $Z$ is \emph{unbounded} in the directions specified by the columns of $\mbf{M}_z$. This allows \eqref{eq:desc_tgrep} to represent symmetrically unbounded sets, such as strips and the entire space $\realset^n$ (Proposition \ref{thm:desc_realspacemgrep}). We use the notation $Z = (\mbf{M}_z,\mbf{G}_z,\mbf{c}_z,\mbf{S}_z,\mbf{A}_z,\mbf{b}_z)\lzon$ for LZs. Moreover, we denote by `$\noarg\!$' an empty argument to this notation. For instance, $(\mbf{G}_z,\mbf{c}_z)\zon = (\noarg,\mbf{G}_z,\mbf{c}_z,\noarg,\noarg,\noarg)\lzon$, and $(\mbf{G}_z,\mbf{c}_z,\mbf{A}_z,\mbf{b}_z)\czon = (\noarg,\mbf{G}_z,\mbf{c}_z,\noarg,\mbf{A}_z,\mbf{b}_z)\lzon$. For convenience, we define $(\mbf{M}_z,\mbf{G}_z,\mbf{c}_z)\lzon \triangleq (\mbf{M}_z,\mbf{G}_z,\mbf{c}_z,\noarg,\noarg,\noarg)\lzon$ for LZs without equality constraints. In addition, we define the set $\mixedball{\mbf{S}_z}{\mbf{A}_z}{\mbf{b}_z} \triangleq \{(\bm{\delta},\bm{\xi}) \in \realset^{n_\delta} \times B_\infty^{n_g}: \mbf{S}_z \bm{\delta} + \mbf{A}_z \bm{\xi} = \mbf{b}_z\}$, which rewrites \eqref{eq:desc_tgrep} as $Z = \{\mbf{c}_z + \mbf{M}_z \bm{\delta} + \mbf{G}_z \bm{\xi} : (\bm{\delta},\bm{\xi}) \in \mixedball{\mbf{S}_z}{\mbf{A}_z}{\mbf{b}_z}\}$.

\begin{figure}[!tb]
	\begin{scriptsize}
		\centering{
			\def\svgwidth{1\columnwidth}			
   			{\scriptsize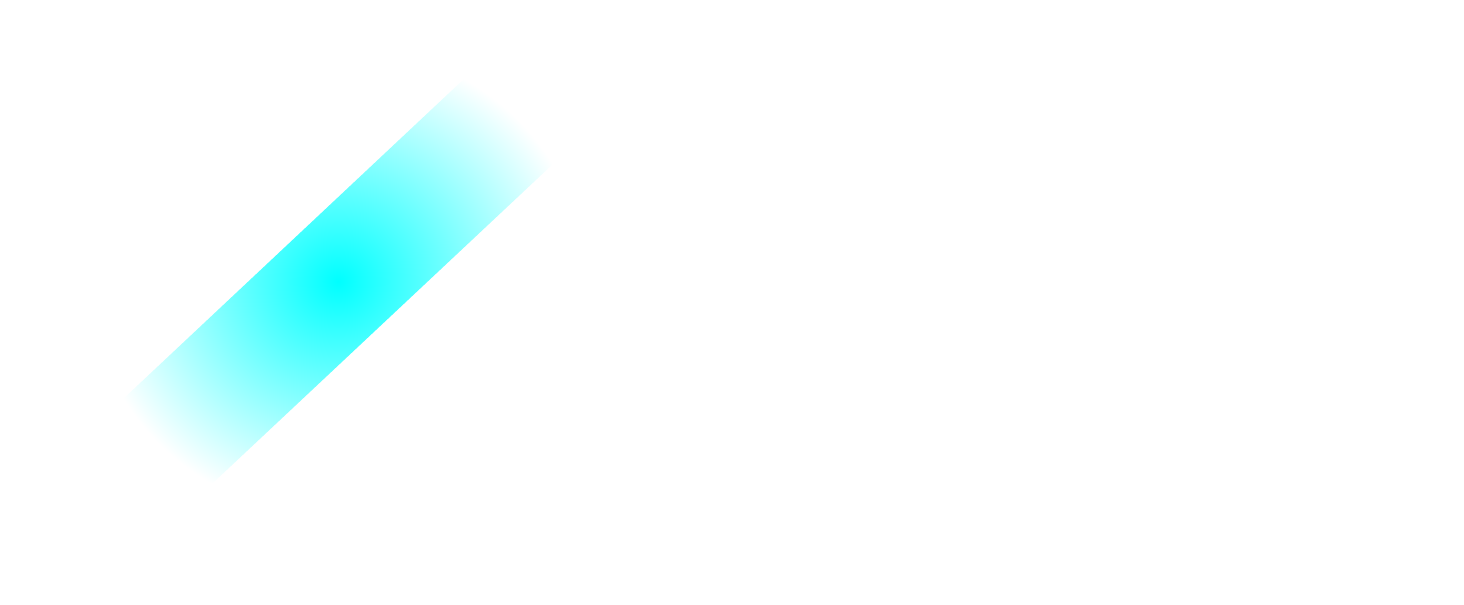}
            \caption{Examples of line zonotopes: a strip $S = \{ \mbf{x} \in \realset^2 : |[-1\,\; 1] \mbf{x} - 1| \leq 0.5\} = \left( \eye{2}, \zeros{2}{1}, \zeros{2}{1}, [-1\,\; 1], -0.5, 1 \right)\lzon$ (left), and the set $( [1 \,\;5 \,\; 3]^T, \mbf{G}_z, \zeros{3}{1}, 0, [-2 \,\; 1  \,\; -1], 2)\lzon$, $\mbf{G}_z \neq \mbf{0}$ (right).}\label{fig:desc_lzonotopeexamples}}	            
	\end{scriptsize}
\end{figure}

Consider line zonotopes $Z, W \subseteq \realset^{n}$, $Y \subseteq \realset^{m}$, i.e., $Z \triangleq (\mbf{M}_z,\mbf{G}_z,\mbf{c}_z,\mbf{S}_z,\mbf{A}_z,\mbf{b}_z)\lzon$, $W \triangleq (\mbf{M}_w,\mbf{G}_w,\mbf{c}_w,\mbf{S}_w,$ $\mbf{A}_w,\mbf{b}_w)\lzon$, $Y \triangleq (\mbf{M}_y,\mbf{G}_y,\mbf{c}_y,\mbf{S}_y,\mbf{A}_y,\mbf{b}_y)\lzon$, and a real matrix $\mbf{R} \in \realset^{m \times n}$. Then, the set operations described in Section \ref{sec:desc_problemformulation} are computed trivially as\footnote{In this work, $\bdiag{\mbf{A},\mbf{B}}$ and $\verti{\mbf{a},\mbf{b}}$ denote $\begin{bmatrix} \mbf{A} & \mbf{0} \\ \mbf{0} & \mbf{B} \end{bmatrix}$ and $\begin{bmatrix} \mbf{a} \\ \mbf{b} \end{bmatrix}$.}
\begin{align}
\mbf{R}Z & = \left( \mbf{R} \mbf{M}_z, \mbf{R} \mbf{G}_z, \mbf{R} \mbf{c}_z, \mbf{S}_z, \mbf{A}_z, \mbf{b}_z \right)\lzon, \label{eq:desc_lzlimage}\\
Z \oplus W & =\left( [\mbf{M}_z \,\; \mbf{M}_w], [\mbf{G}_z \,\; \mbf{G}_w],  \mbf{c}_z + \mbf{c}_w, \right. \nonumber \\ & \quad\left. \bdiag{\mbf{S}_z,\mbf{S}_w}, \bdiag{\mbf{A}_z,\mbf{A}_w}, \verti{\mbf{b}_z,\mbf{b}_w}\right)\lzon, \label{eq:desc_lzmsum}\\
Z \cap_{\mbf{R}} Y & = \left( [\mbf{M}_z \,\; \bm{0}], [\mbf{G}_z \,\; \bm{0}], \mbf{c}_z, \right. \nonumber\\ & \left. \begin{bmatrix} \multicolumn{2}{c}{\bdiag{\mbf{S}_z,\mbf{S}_y}} \\ \mbf{R} \mbf{M}_z & -\mbf{M}_y \end{bmatrix}, \begin{bmatrix}\multicolumn{2}{c}{\bdiag{\mbf{A}_z,\mbf{A}_y}} \\ \mbf{R} \mbf{G}_z & -\mbf{G}_y \end{bmatrix}, \begin{bmatrix} \verti{\mbf{b}_z,\mbf{b}_y} \\ \mbf{c}_y - \mbf{R} \mbf{c}_z \end{bmatrix} \right)\lzon\!\!\!\!\!. \label{eq:desc_lzintersection}\\
Z \times W & = \left( \bdiag{\mbf{M}_z, \mbf{M}_w},\bdiag{\mbf{G}_z,\mbf{G}_w}, \verti{\mbf{c}_z,\mbf{c}_w}, \right. \nonumber \\ & \quad \left. \bdiag{\mbf{S}_z,\mbf{S}_w}, \bdiag{\mbf{A}_z,\mbf{A}_w}, \verti{\mbf{b}_z,\mbf{b}_w} \right)\lzon, \label{eq:desc_lzcartprod}
\end{align}
The derivation of \eqref{eq:desc_lzlimage}--\eqref{eq:desc_lzintersection} follows the same approach as the CZ case, which is detailed in \cite{Scott2016}, while \eqref{eq:desc_lzcartprod} is straightforward. Additionally, note that a CZ is a special case of a LZ (i.e., without lines), and, consequently, every zonotope is also a LZ. The following proposition demonstrates that other classes of sets can also be represented using LZs. Figure \ref{fig:desc_lzonotopeexamples} provides examples of unbounded sets described by LZs.

\begin{proposition} \rm \label{thm:desc_realspacemgrep}
	The set $\realset^n$, strips, and hyperplanes are LZs.
\end{proposition}
\proof Let $\mbf{M}_\text{R} \in \realsetmat{n}{n_\delta}$ be a full row rank matrix, and let $\mbf{c}_\text{R} \in \realset^n$. Since $\text{rank}(\mbf{M}_\text{R}) = n$, for every $\mbf{z} \in \realset^n$ there exists at least one $\bm{\delta} \in \realset^{n_\delta}$ such that $\mbf{z} = \mbf{c}_\text{R} + \mbf{M}_\text{R} \bm{\delta}$. Therefore, $\realset^n \subseteq \left( \mbf{M}_\text{R}, \noarg, \mbf{c}_\text{R}\right)\lzon$. On the other side, choose one $\bm{\delta} \in \realset^{n_\delta}$, and define $\mbf{r} \triangleq \mbf{c}_\text{R} + \mbf{M}_\text{R} \bm{\delta}$. Since $\text{rank}(\mbf{M}_\text{R}) = n$, there must exist $\mbf{z} \in \realset^n$ such that $\mbf{z} = \mbf{r}$. Therefore, $\left( \mbf{M}_\text{R}, \noarg, \mbf{c}_\text{R} \right)\lzon \subseteq \realset^n$, which implies $\realset^n = \left( \mbf{M}_\text{R}, \noarg, \mbf{c}_\text{R} \right)\lzon$. Consider a strip $S = \{ \mbf{x} \in \realset^n : |\bm{\rho}_s^T \mbf{x} - d_s| \leq \sigma_s\}$, with $\bm{\rho}_s \in \realset^n$, $d_s,\sigma_s \in \realset$, $\sigma_s \geq 0$. Note that an equivalent definition of $S$ is $\{ \mbf{x} \in \realset^n : \bm{\rho}_s^T \mbf{x} \in [-\sigma_s + d_s, \sigma_s + d_s]\}$. Therefore, by writing the interval $[-\sigma_s + d_s, \sigma_s + d_s]$ in G-rep as $(\sigma_s, d_s)\zon$, or equivalently in CLG-rep as $(\emptyarg,\sigma_s,d_s)\lzon$, and $\realset^n = (\eye{n},\emptyarg,\zeros{n}{1})\lzon$, we have that $S$ can be described in CLG-rep as $S = \realset^n \cap_{\bm{\rho}^T} (\sigma_s, d_s)\zon = ( \eye{n}, \zeros{n}{1}, \zeros{n}{1}, \bm{\rho}_s^T, -\sigma_s, d_s )\lzon$. Finally, a hyperplane is a particular case of a strip given by $\{ \mbf{x} \in \realset^n : |\bm{\rho}_s^T \mbf{x} - d_s| \leq 0\} = \{ \mbf{x} \in \realset^n : \bm{\rho}_s^T \mbf{x} = d_s\}$ (i.e., a degenerated strip), which concludes the proof.
\qed

\begin{remark} \rm \label{rem:realset}
    Without loss of generality, in the remainder of the paper we choose (\eye{n},\emptyarg,\zeros{n}{1})\lzon \ to describe $\realset^n$ in CLG-rep.
\end{remark}

\begin{remark} \rm \label{rem:ccgs}
 LZs are a special case of the CCGs in \cite{Silvestre2022}. Unlike general CCGs, which can also represent unbounded sets, LZs offer the advantage of efficient polynomial-time complexity reduction methods (see Section \ref{sec:desc_lzcomplexityreduction}).
\end{remark}

\subsection{Complexity reduction of line zonotopes} \label{sec:desc_lzcomplexityreduction}

Similar to zonotopes and constrained zonotopes, the set operations in \eqref{eq:desc_lzlimage}--\eqref{eq:desc_lzcartprod} lead to a linear increase in the number of lines, generators, and constraints in the CLG-rep \eqref{eq:desc_tgrep}. When these operations are performed iteratively, as it is common in set-based state estimation, the complexity of the set may increase indefinitely. This section introduces methods for reducing the complexity of LZs by outer-approximating them with simpler ones that have fewer lines, generators, and constraints.

\subsubsection{Line elimination}

The proposed procedure for eliminating lines closely follows the constraint elimination method used for CZs, as explained below. Let $Z = (\mbf{M}_z,\mbf{G}_z,\mbf{c}_z,\mbf{S}_z,\mbf{A}_z,\mbf{b}_z)\lzon \subseteq \realset^n$ be a line zonotope with $n_g$ generators, $n_\delta$ lines, and $n_c$ constraints. 
It will be shown that the maximum number of lines that can be eliminated is equal to the row rank of $\mbf{S}_z$. Additionally, this procedure does not introduce any conservatism. The procedure is analogous to the constraint elimination method in \cite{Scott2016}, with the key difference that the variable selected for elimination is a line, rather than a generator. This is done by solving the constraints $\mbf{S}_z \bm{\delta} + \mbf{A}_z \bm{\xi} = \mbf{b}_z$ for a subset of the $\bm{\delta}$ variables, with size equal to the row rank of $\mbf{S}_z$, and substituting these in $\mbf{c}_z + \mbf{M}_z \bm{\delta} + \mbf{G}_z \bm{\xi}$ in \eqref{eq:desc_tgrep}. This procedure simultaneously eliminates an equal number of constraints from $Z$. However, in contrast to eliminating a generator, $\bm{\delta}$ is unbounded, hence no information is lost, resulting in an equivalent set. Furthermore, if all the lines of $Z$ can be eliminated through this procedure, then $Z$ becomes a CZ. Algorithm \ref{alg:desc_lzugennelim} summarizes the proposed procedure, which has complexity $O((n + n_c)(n_\delta + n_g))$ for each line eliminated.

\begin{algorithm}[!tb]
	\caption{Elimination of one line from \eqref{eq:desc_tgrep}.}
	\label{alg:desc_lzugennelim}
	\small
	\begin{algorithmic}[1]
		\State Let $Z \triangleq (\mbf{M},\mbf{G},\mbf{c},\mbf{S},\mbf{A},\mbf{b})\lzon$. Choose $i,j$ such that  $S_{i,j} \neq 0$.
        \State Solve the $i$-th constraint for $\delta_j$:
\vspace{-0.2cm}
\[
\delta_j = S_{i,j}^{-1} \left( b_i - \sum_{m \neq j} S_{i,m} \delta_m - \sum_{\ell} A_{i, \ell} \xi_\ell \right).
\]
\vspace{-0.2cm}
		\State Substitute the solution for $\delta_j$ into $\mbf{S} \bm{\delta} + \mbf{A} \bm{\xi} = \mbf{b}$ and $\mbf{c} + \mbf{M} \bm{\delta} + \mbf{G} \bm{\xi}$.
	\end{algorithmic}
	\normalsize
\end{algorithm}	

\subsubsection{Constraint elimination and generator reduction}\label{sec_312}

Let $Z = (\mbf{M}_z,\mbf{G}_z,\mbf{c}_z,\mbf{S}_z,\mbf{A}_z,\mbf{b}_z)\lzon \subseteq \realset^n$. Constraint elimination in LZs can be performed in the same way as for CZs, using the method described in \cite{Scott2016}. However, since the removal of a line is not conservative, if $k_\delta\triangleq\text{rank}(\mbf{S}_z) > 0$ then $k_\delta$ lines are selected for elimination along with $k_\delta$ constraints, before any generators are removed from $Z$. The complexity of eliminating a single constraint using one generator is $O((n_g +n_c)^3)$.
It is important to note that it is is always possible to rewrite the LZ constraints  $\mbf{S}_z  \bm{\delta} + \mbf{A}_z \bm{\xi} = \mbf{b}_z$
as  
$\begin{bmatrix} {\mbf{I}_{k_\delta\times k_\delta}} & \mbf{S}^+_z & \mbf{A}^+_z \\ \mbf{0}_{n_c-k_\delta\times k_\delta} & \mbf{0}_{n_c-k_\delta\times k_\delta} & \mbf{A}^-_z\end{bmatrix} 
\begin{bmatrix}
\bm{\delta}\\ \bm{\xi} \end{bmatrix}=\begin{bmatrix}\bm{b}^+_z\\\bm{b}^-_z\end{bmatrix} 
$ with suitable matrices $\mbf{S}^+_z\in \mathbb{R}^{k_\delta\times n_d-k_\delta}$, $\mbf{A}^+_z \in \mathbb{R}^{k_\delta\times n_g}$, $\mbf{A}^-_z\in \mathbb{R}^{n_c-k_\delta\times n_g}$, $\bm{b}^+_z\in \mathbb{R}^{k_\delta}$ $\bm{b}^-_z\in \mathbb{R}^{nc-k_\delta}$ using, e.g., Gauss-Jordan elimination.  Examining this expression, we observe that eliminating the $k_\delta$ lines results in constraints that involve only the generators, i.e.,  $\mbf{A}_z^-\bm{\xi}=\bm{b}_z^-$. Let $Z^- = (\mbf{M}_z^-,\mbf{G}_z^-,\mbf{c}_z^-,\mbf{0},\mbf{A}_z^-,\mbf{b}_z^-)\lzon \subseteq \realset^n$ be the equivalent LZ obtained after removing $k_\delta = \text{rank}(\mbf{S}_z)$ constraints and lines from $Z$. Then, the resulting LZ $Z^-$ can be decoupled as $Z^- = \mbf{c}_z^- \oplus \mbf{M}_z^- \realset^{n_\delta^-} \oplus \mbf{G}_z^- B_\infty(\mbf{A}_z^-,\mbf{b}_z^-)$. %
At this point, generator reduction of LZs can be carried out in the same manner as for CZs, as described in \cite{Scott2016}, by first lifting  $\mbf{G}_z^- B_\infty(\mbf{A}_z^-,\mbf{b}_z^-)$, and then applying a zonotope order reduction method. The resulting line zonotope is given by $\bar{Z} = \mbf{c}_z^- \oplus \mbf{M}_z^- \realset^{n_\delta} \oplus \bar{\mbf{G}}_z B_\infty(\bar{\mbf{A}}_z,\mbf{b}_z^-) \supseteq Z^-$, with $\bar{\mbf{G}}_z$ and $\bar{\mbf{A}}_z$ being the resulting matrices provided by the zonotope generator reduction method. The reduction by $k_g$ generators has complexity $O(n^2n_g +k_gn_gn)$.

\subsection{Some examples using line zonotopes} \label{app:examples}

\subsubsection{Intersection of a zonotope and a strip}

Since both zonotopes and strips are line zonotopes, the intersection of a zonotope and a strip can be computed trivially and efficiently in CLG-rep, using the generalized intersection \eqref{eq:desc_lzintersection} as illustrated below. %
This operation often appears in the update step of set-based state estimation using zonotopes. Consider the zonotope \citep{Bravo2006}
\begin{equation*}
Z = \left(\begin{bmatrix} 0.2812 & 0.1968 & 0.4235 \\ 0.0186 & -0.2063 & -0.2267 \end{bmatrix}, \begin{bmatrix} 0 \\ 0 \end{bmatrix} \right)\zon,
\end{equation*}
and the strip $S = \{ \mbf{x} \in \realset^2 : |\bm{\rho}_s^T \mbf{x} - d_s| \leq \sigma_s\}$, with $\bm{\rho}_s = [1 \; -1]^T$, $d_s = 1$, and $\sigma_s = 0.1$. The intersection $Z \cap S$ is computed by writing $S$ in CLG-rep, then using \eqref{eq:desc_lzintersection}. The resulting set is
\begin{equation*}
\begin{aligned}
Z \cap S = & \left(\begin{bmatrix} 0 & 0 \\ 0 & 0 \end{bmatrix}, \begin{bmatrix} 0.2812 & 0.1968 & 0.4235 & 0 \\ 0.0186 & -0.2063 & -0.2267 & 0 \end{bmatrix}, \begin{bmatrix} 0 \\ 0 \end{bmatrix}, \right. \\ & \!\!\!\!\!\!\! \left. \begin{bmatrix} 1 & -1 \\ -1 & 0 \\ 0 & -1 \end{bmatrix}, \begin{bmatrix} 0 & 0 & 0 & -0.1 \\ 0.2812 & 0.1968 & 0.4235 & 0 \\ 0.0186 & -0.2063 & -0.2267 & 0 \end{bmatrix}, \begin{bmatrix} 1 \\ 0 \\ 0 \end{bmatrix} \right)\lzon.
\end{aligned}
\end{equation*} 

\begin{figure}[!htb]
	\centering{
		\def\svgwidth{0.95\columnwidth}
		{\scriptsize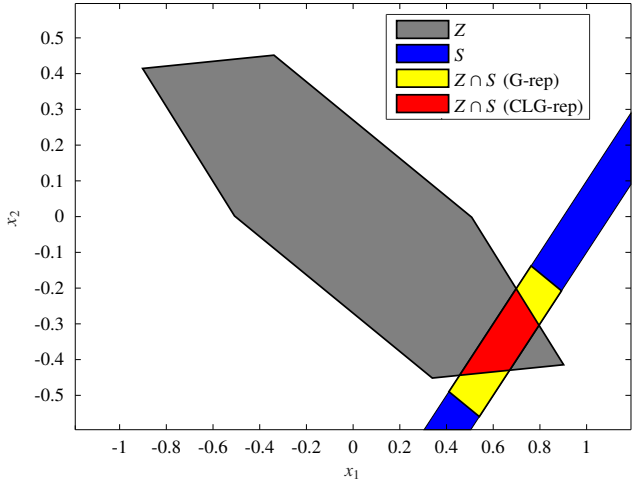}
		\caption{The zonotope $Z$ (gray), the strip $S$ (blue), the zonotope obtained using the conservative intersection method in \cite{Bravo2006} (yellow), and the line zonotope $Z \cap S$ computed in CLG-rep (red)}\label{fig:desc_intersectionzonstrip}}
\end{figure}

Figure \ref{fig:desc_intersectionzonstrip} shows the zonotope $Z$ (gray) and strip $S$ (blue), as well as the zonotope obtained using the intersection method in \cite{Bravo2006} (yellow) and the line zonotope $Z \cap S$ computed in CLG-rep (red). The latter is obtained using Proposition \ref{thm:desc_realspacemgrep} and \eqref{eq:desc_lzintersection}, and as it can be noticed, it corresponds to the exact intersection of the zonotope $Z$ and the strip $S$. This result can also be obtained using constrained zonotopes through generalized intersection, unlike the operations presented next.

\subsubsection{Set operations with strips} \label{sec:stripoperations}

\begin{figure}[!htb]
	\centering{
		\def\svgwidth{1\columnwidth}
		{\scriptsize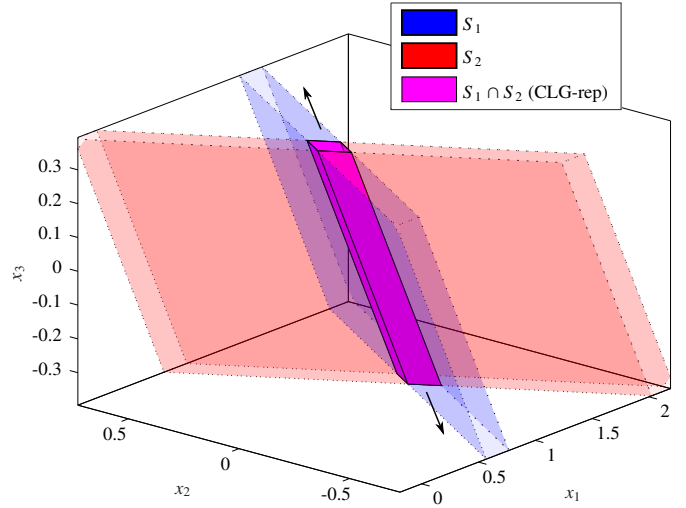}
		\caption{The strips $S_1$ (blue) and $S_2$ (red), and the intersection $S_1 \cap S_2$ computed computed in CLG-rep (magenta).}\label{fig:desc_intersectionstrips}}
\end{figure}

The operations \eqref{eq:desc_lzlimage}--\eqref{eq:desc_lzcartprod} are computed efficiently for strips in CLG-rep, which may be useful, for instance, in state estimation when the initial set is unknown. In particular, let $S_i \subset \realset^n$ be strips, with $i \in \{1 \ldots,p\}$. The intersection $\bigcap_{i=1}^p S_i$ is not bounded in the general case, and therefore, it cannot be computed using constrained zonotopes. Moreover, for $p < n$, this intersection is not a strip in general. On the other side, since $S_i$ can be written in CLG-rep using Proposition \ref{thm:desc_realspacemgrep}, the set $\bigcap_{i=1}^p S_i$ is then a line zonotope and can be computed using \eqref{eq:desc_lzintersection}. %
Let $S_1 \triangleq \{ \mbf{x} \in \realset^3 : |\bm{\rho}_1^T \mbf{x} - d_1| \leq \sigma_1\}$ and $S_2 \triangleq \{ \mbf{x} \in \realset^3 : |\bm{\rho}_2^T \mbf{x} - d_2| \leq \sigma_2\}$, with $\bm{\rho}_1 = [1 \,\; -1 \,\; 1]^T$, $\bm{\rho}_2 = [1 \,\; 1 \,\; 1]^T$, $d_1 = 1$, $d_2 = 1$, $\sigma_1 = 0.1$, and $\sigma_2 = 0.1$. Figure \ref{fig:desc_intersectionstrips} shows the strips $S_1$ (blue) and $S_2$ (red), as well as the intersection $S_1 \cap S_2$ computed in CLG-rep (magenta). It is worth highlighting that the intersection in this case cannot be expressed as a constrained zonotope or a strip, while it can be described exactly using the CLG-rep \eqref{eq:desc_tgrep}.

\subsubsection{Enclosing the domain of $p$ linear mappings}
Consider $p$ sets $Z_i \subset \realset^{n_i}$ and $p$ real matrices $\mbf{R}_i \in \realsetmat{n_i}{n}$, with $i \in \{1,\ldots,p\}$. Define the operator
\begin{equation} \label{eq:desc_twointersection}
\bigcap_{\mbf{R}_i}^{\{1,p\}} Z_i \triangleq \{ \mbf{x} \in \realset^n: \mbf{R}_1 \mbf{x} \in ~Z_1,\mbf{R}_2 \mbf{x} \in Z_2,\ldots, \mbf{R}_p \mbf{x} \in Z_p\}.
\end{equation}

Note that the generalized intersection \eqref{eq:desc_lzintersection} is a particular case of \eqref{eq:desc_twointersection} in which $p = 2$, $\mbf{R}_1 = \eyenoarg$ and $\mbf{R}_2 = \mbf{R}$. Moreover, \eqref{eq:desc_twointersection} is also a generalization of the intersection of strips described in Section \ref{sec:stripoperations}. If the sets $Z_i$ are in CLG-rep, the operator \eqref{eq:desc_twointersection} can be computed as follows. Let $\bigcap_{\mbf{R}_i}^{\{1,p\}} Z_i$ be rewritten as $\bigcap_{\mbf{R}_i}^{\{1,p\}} Z_i  = \realset^n \cap_{\tilde{\mbf{R}}} \tilde{Z}$, where $\tilde{\mbf{R}} \triangleq[\mbf{R}_1^T \; \cdots \; \mbf{R}_p^T]^T$, and $\tilde{\mbf{Z}} \triangleq Z_1 \times \ldots \times Z_p$. Moreover, let $Z_i = \{\mbf{M}_i, \mbf{G}_i, \mbf{c}_i, \mbf{S}_i, \mbf{A}_i, \mbf{b}_i\}$, with $n_{\delta_i}$ lines, and $n_{g_i}$ generators. Additionally, $\tilde{\mbf{M}} \triangleq \text{bdiag}(\mbf{M}_1,\ldots,\mbf{M}_p)$, $\tilde{\mbf{G}} \triangleq \text{bdiag}(\mbf{G}_1,\ldots,\mbf{G}_p)$, $\tilde{\mbf{c}} \triangleq [\mbf{c}_1^T \; \cdots \; \mbf{c}_p^T]^T$, $\tilde{\mbf{S}} \triangleq \text{bdiag}(\mbf{S}_1,\ldots,\mbf{S}_p)$, $\tilde{\mbf{A}} \triangleq \text{bdiag}(\mbf{A}_1,\ldots,\mbf{A}_p)$, $\tilde{\mbf{b}} \triangleq [\mbf{b}_1^T \; \cdots \; \mbf{b}_p^T]^T$, $\tilde{n}_\delta \triangleq \sum_i^p n_{\delta_i}$, $\tilde{n}_g \triangleq \sum_i^p n_{g_i}$, and let $\realset^n = (\eye{n}, \emptyarg, \zeros{n}{1})\lzon$. Then, using \eqref{eq:desc_lzintersection} leads to
\begin{equation*}
\bigcap_{\mbf{R}_i}^{\{1,p\}} Z_i = \left( [\eye{n} \,\; \zeros{n}{\tilde{n}_{\delta}}], \zeros{n}{\tilde{n}_{g}}, \zeros{n}{1}, \begin{bmatrix} \mbf{0} & \tilde{\mbf{S}} \\ \tilde{\mbf{R}} & -\tilde{\mbf{M}} \end{bmatrix}, \begin{bmatrix} \tilde{\mbf{A}} \\ - \tilde{\mbf{G}}\end{bmatrix}, \begin{bmatrix} \tilde{\mbf{b}} \\ \tilde{\mbf{c}} \end{bmatrix} \right)\lzon.
\end{equation*}

\section{Set-based state estimation of LDS using LZs}  \label{sec:desc_estimationLZ}

In state estimation, $i \in \modelset$ is assumed to be known and constant. Thus, without loss of generality, we omit the superscripts $i$ from \eqref{eq:desc_system}. Given $X_0$, input $\mbf{u}_0$, and output $\mbf{y}_0$, let
\begin{equation} \label{eq:desc_initialsetlz}
\hat{X}_0 \triangleq \{ \mbf{x} \in X_0 : \mbf{C} \mbf{x} + \mbf{D} \mbf{u}_{0} + \mbf{D}_v \mbf{v} = \mbf{y}_0, \, \mbf{v} \in V \}.
\end{equation}
Let $\mbf{E} = \mbf{U} \mbf{\Sigma} \mbf{V}^T$ be the result of a singular value decomposition, where $\mbf{U}$ and $\mbf{V}$ are invertible by construction. Since $\mbf{E}$ is square, then $\mbf{\Sigma}$ is also square. Let $\mbf{\Sigma}$ be arranged as $\mbf{\Sigma} = \bdiag{\tilde{\mbf{\Sigma}}, \mbf{0}}$, where $\tilde{\mbf{\Sigma}} \in \realsetmat{n_z}{n_z}$ is diagonal containing the $n_z = \text{rank}(\mbf{E})$ nonzero singular values of $\mbf{E}$. Moreover, given $\mbf{T} = \inv{(\mbf{V}^T)}$, let $\mbf{z}_k = (\tilde{\mbf{z}}_k,\check{\mbf{z}}_k) = \inv{\mbf{T}} \mbf{x}_k, \; \tilde{\mbf{z}}_k \in \realset^{n_z}, \check{\mbf{z}}_k \in \realset^{n-n_z}$,
\begin{equation}\label{eq:desc_SVDmatriceslz} 
\begin{aligned}
\begin{bmatrix} \tilde{\mbf{A}} \\ \check{\mbf{A}} \end{bmatrix} & = \begin{bmatrix} \tilde{\mbf{\Sigma}}^{-1} & \mbf{0} \\ \mbf{0} & \eyenoarg \end{bmatrix} \inv{\mbf{U}} \mbf{A} \mbf{T}, \\
\begin{bmatrix} \tilde{\mbf{B}} \\ \check{\mbf{B}} \end{bmatrix} & = \begin{bmatrix} \tilde{\mbf{\Sigma}}^{-1} & \mbf{0} \\ \mbf{0} & \eyenoarg \end{bmatrix} \inv{\mbf{U}} \mbf{B}, \begin{bmatrix} \tilde{\mbf{B}}_w \\ \check{\mbf{B}}_w \end{bmatrix} = \begin{bmatrix} \tilde{\mbf{\Sigma}}^{-1} & \mbf{0} \\ \mbf{0} & \eyenoarg \end{bmatrix} \inv{\mbf{U}} \mbf{B}_w,  \\
\end{aligned}
\end{equation}
with $\tilde{\mbf{A}} \in \realsetmat{n_z}{n}$, $\tilde{\mbf{B}} \in \realsetmat{n_z}{n_u}$, $\tilde{\mbf{B}}_w \in \realsetmat{n_z}{n_w}$. Then, as in \cite{Rego2020b}, system \eqref{eq:desc_system} can be rewritten as
\begin{subequations} \label{eq:desc_systemSVDlz}
	\begin{align} 
	\tilde{\mbf{z}}_k & = \tilde{\mbf{A}} \mbf{z}_{k-1} + \tilde{\mbf{B}} \mbf{u}_{k-1} + \tilde{\mbf{B}}_w \mbf{w}_{k-1},  \label{eq:desc_systemSVDdynamicslz} \\
	\mbf{0} & = \check{\mbf{A}} \mbf{z}_{k} + \check{\mbf{B}} \mbf{u}_{k} + \check{\mbf{B}}_w \mbf{w}_{k}, \label{eq:desc_systemSVDconstraintslz} \\
	\mbf{y}_k & = \mbf{C} \mbf{T} \mbf{z}_k + \mbf{D} \mbf{u}_{k} + \mbf{D}_v \mbf{v}_{k}. \label{eq:desc_systemSVDoutputlz}
	\end{align}
\end{subequations}

Starting with the initialization provided by \eqref{eq:desc_initialsetlz}, and setting $\hat{Z}_0 \triangleq \{\mbf{T}^{-1} \mbf{x}_0 : \mbf{x}_0 \in \hat{X}_0, ~\check{\mbf{A}} \mbf{T}^{-1} \mbf{x}_0 + \check{\mbf{B}} \mbf{u}_0 + \check{\mbf{B}}_w \mbf{w}_0 = \mbf{0}\}$, we then define, for $k \geq 1$, the \emph{prediction} and \emph{update steps} as the computation of the enclosures $\bar{Z}_k$ and $\hat{Z}_k$, which satisfy
\begin{align}
\bar{Z}_k & \hspace{-0.05cm}\supseteq \hspace{-0.05cm}\{\mbf{z}_{k}\hspace{-0.05cm}=\hspace{-0.05cm}(\tilde{\mbf{A}} \mbf{z}_{k-1}\hspace{-0.1cm} +\hspace{-0.05cm} \tilde{\mbf{B}} \mbf{u}_{k-1} \hspace{-0.05cm}+\hspace{-0.05cm} \tilde{\mbf{B}}_w \mbf{w}_{k-1}, \check{\mbf{z}}_k)\hspace{-0.05cm}:\hspace{-0.05cm} \mbf{0} = \hspace{-0.05cm}\check{\mbf{A}} \mbf{z}_{k} \hspace{-0.05cm}+\hspace{-0.05cm} \check{\mbf{B}} \mbf{u}_{k} \hspace{-0.05cm}+\hspace{-0.05cm} \check{\mbf{B}}_w \mbf{w}_{k}, \nonumber \\ & \qquad \quad  (\mbf{z}_{k-1},\check{\mbf{z}}_k,\mbf{w}_{k-1},\mbf{w}_k) \in \hat{Z}_{k-1} \times \realset^{n-n_z} \times W \times W\}, \label{eq:predictionlds} \\
\hat{Z}_k & \supseteq \{\mbf{z}_k \in \bar{Z}_k : \mbf{y}_k = \mbf{C} \mbf{T} \mbf{z}_k + \mbf{D} \mbf{u}_{k} + \mbf{D}_v \mbf{v}_{k}, \mbf{v}_k \in V\}, \label{eq:updatelds}
\end{align}
with the estimated enclosure $\hat{X}_k$ given by $\hat{X}_k \triangleq \{\mbf{T} \mbf{z}_k : \mbf{z}_k \in \hat{Z}_k\}$.

\subsection{Proposed method}

This section introduces a novel method for set-based state estimation of the system \eqref{eq:desc_system} using line zonotopes. The capability to describe unbounded sets enables state estimation of the LDS \eqref{eq:desc_system} via the CLG-rep offering two key advantages: (i) it eliminates the need for a bounded initial set satisfying $\mbf{x}_0 \in X_0$, where in this case, $X_0 = \realset^n$; and (ii) it removes the requirement for the existence of an admissible set as stipulated in Assumption 1 of \cite{Rego2020b}. 

Let $W = (\mbf{M}_w, \mbf{G}_w, \mbf{c}_w, \mbf{S}_w, \mbf{A}_w, \mbf{b}_w)\lzon$, and $\hat{X}_{0}$ given by \eqref{eq:desc_initialsetlz}. From \eqref{eq:desc_systemSVDconstraintslz}, the state $\mbf{z}_0$ must satisfy $\check{\mbf{A}} \mbf{z}_0 + \check{\mbf{B}} \mbf{u}_0 + \check{\mbf{B}}_w \mbf{w}_0 = \mbf{0}$. This constraint is incorporated in the CLG-rep of $\hat{Z}_0$ as follows. Let $\inv{\mbf{T}} \hat{X}_{0} \triangleq (\mbf{M}_0, \mbf{G}_0,\mbf{c}_0,\mbf{S}_0,\mbf{A}_0,\mbf{b}_0)\lzon$. Then, $\hat{Z}_0 \triangleq (\hat{\mbf{M}}_0,\hat{\mbf{G}}_0, $ $\hat{\mbf{c}}_0,\hat{\mbf{S}}_0,\hat{\mbf{A}}_0,\hat{\mbf{b}}_0)\lzon$, with $\hat{\mbf{M}}_0 = [\mbf{M}_0 \,\; \mbf{0}] ,\hat{\mbf{G}}_0 = [\mbf{G}_0 \,\; \mbf{0}]$, $\hat{\mbf{c}}_0 = \mbf{c}_0$,
\begin{equation*}
\begin{aligned}
\hat{\mbf{S}}_0 & = \begin{bmatrix} \multicolumn{2}{c}{\bdiag{\mbf{S}_0, \mbf{S}_w)}} \\ \check{\mbf{A}} \mbf{M}_0 & \check{\mbf{B}}_w \mbf{M}_w \end{bmatrix},\; \hat{\mbf{A}}_0 = \begin{bmatrix} \multicolumn{2}{c}{\bdiag{\mbf{A}_0,\mbf{A}_w}} \\ \check{\mbf{A}} \mbf{G}_0 & \check{\mbf{B}}_w \mbf{G}_w \end{bmatrix}, \\
\hat{\mbf{b}}_0 & = \begin{bmatrix} \verti{\mbf{b}_0,\mbf{b}_w} \\ -\check{\mbf{A}} \mbf{c}_0 - \check{\mbf{B}}_w \mbf{c}_w - \check{\mbf{B}} \mbf{u}_{0} \end{bmatrix}.
\end{aligned}
\end{equation*}
Note that the extra columns in $\hat{\mbf{G}}_0$ and $\hat{\mbf{A}}_0$ come from $\mbf{w}_0 \in W$. It is also noteworthy that, at each $k \geq 0$, the variables $\tilde{\mbf{z}}_k$ are fully determined by \eqref{eq:desc_systemSVDdynamicslz}, while $\check{\mbf{z}}_k$ are obtained a posteriori by \eqref{eq:desc_systemSVDconstraintslz}. %
An effective enclosure of the prediction step for the descriptor system \eqref{eq:desc_systemSVDlz} can be obtained in CLG-rep as follows. 

\begin{lemma} \label{lem:desc_predictionlz} \rm
    Consider the transformed state-space \eqref{eq:desc_systemSVDlz}. Let $\mbf{z}_{k-1} \in \hat{Z}_{k-1} \triangleq (\hat{\mbf{M}}_{k-1}, \hat{\mbf{G}}_{k-1},$ $ \hat{\mbf{c}}_{k-1}$, $\hat{\mbf{S}}_{k-1}$, $\hat{\mbf{A}}_{k-1}, \hat{\mbf{b}}_{k-1})\lzon$, and $\mbf{w}_{k-1}, \mbf{w}_k \in W \triangleq (\mbf{M}_w, \mbf{G}_w, \mbf{c}_w, \mbf{S}_w, \mbf{A}_w, \mbf{b}_w)\lzon$, for all $k \geq 1$. Then $\mbf{z}_k \in \bar{Z}_k \triangleq (\bar{\mbf{M}}_{k}, \bar{\mbf{G}}_{k}, \bar{\mbf{c}}_{k}$, $\bar{\mbf{S}}_{k}$, $\bar{\mbf{A}}_{k}, \bar{\mbf{b}}_{k})\lzon$, with
	\begin{align*}
	\bar{\mbf{M}}_k & = \begin{bmatrix} \tilde{\mbf{A}} \hat{\mbf{M}}_{k-1} & \tilde{\mbf{B}}_w \mbf{M}_{w} & \bm{0} & \bm{0} \\ \mbf{0} & \mbf{0} & \bm{0} & \eye{n-n_z} \end{bmatrix}, \bar{\mbf{G}}_k {=} \begin{bmatrix} \tilde{\mbf{A}} \hat{\mbf{G}}_{k-1} & \tilde{\mbf{B}}_w \mbf{G}_w & \mbf{0} \\
	\mbf{0} & \mbf{0} & \mbf{0}\end{bmatrix}, \\
	\bar{\mbf{c}}_k & = \begin{bmatrix} \tilde{\mbf{A}} \hat{\mbf{c}}_{k-1} + \tilde{\mbf{B}} \mbf{u}_{k-1} + \tilde{\mbf{B}}_w \mbf{c}_w \\ \zeros{(n-n_z)}{1} \end{bmatrix}\!, \\  
	\bar{\mbf{S}}_k & = \left[\begin{array}{ccc|c} \multicolumn{3}{c|}{\bdiag{\hat{\mbf{S}}_{k-1}, \mbf{S}_w, \mbf{S}_w)}} & \bm{0} \\ \hline \check{\mbf{A}} \begin{bmatrix} \tilde{\mbf{A}} \hat{\mbf{M}}_{k-1} \\ \bm{0} \end{bmatrix} & \check{\mbf{A}} \begin{bmatrix} \tilde{\mbf{B}}_w \mbf{M}_w \\ \bm{0} \end{bmatrix} & \mbf{0} & \check{\mbf{A}} \begin{bmatrix} \bm{0} \\ \eye{n-n_z} \end{bmatrix} \end{array}\right], \\
	\bar{\mbf{A}}_k & = \left[\begin{array}{ccc} \multicolumn{3}{c}{\bdiag{\hat{\mbf{A}}_{k-1},\mbf{A}_w,\mbf{A}_w)}} \\ \hline
	\check{\mbf{A}} \begin{bmatrix} \tilde{\mbf{A}} \hat{\mbf{G}}_{k-1}  \\ \mbf{0} \end{bmatrix} & \check{\mbf{A}} \begin{bmatrix} \tilde{\mbf{B}}_w \mbf{G}_w \\ \mbf{0} \end{bmatrix} & \check{\mbf{B}}_w \mbf{G}_w \end{array}\right]\!, \\
	\bar{\mbf{b}}_k & = \begin{bmatrix} \verti{\hat{\mbf{b}}_{k-1},\mbf{b}_w,\mbf{b}_w} \\ -\check{\mbf{A}} \begin{bmatrix} \tilde{\mbf{A}} \hat{\mbf{c}}_{k-1} + \tilde{\mbf{B}} \mbf{u}_{k-1} + \tilde{\mbf{B}}_w \mbf{c}_w \\ \zeros{n-n_z}{1} \end{bmatrix} - \check{\mbf{B}} \mbf{u}_k - \check{\mbf{B}}_w \mbf{c}_w \end{bmatrix}.                                
	\end{align*}
\end{lemma}
\proof 
Since by assumption $(\mbf{z}_{k-1},\mbf{w}_{k-1},\mbf{w}_k) \in \hat{Z}_{k-1} \times W \times W$, there exists $((\bm{\delta}_{k-1},\bm{\xi}_{k-1}), (\bm{\vartheta}_{k-1},\bm{\varphi}_{k-1}), (\bm{\vartheta}_{k},\bm{\varphi}_{k})) \in \mixedball{\hat{\mbf{S}}_{k-1}}{\hat{\mbf{A}}_{k-1}}{\hat{\mbf{b}}_{k-1}} \times \mixedball{\mbf{S}_w}{\mbf{A}_w}{\mbf{b}_w} \times \mixedball{\mbf{S}_w}{\mbf{A}_w}{\mbf{b}_w}$ such that $\mbf{z}_{k-1} = \hat{\mbf{c}}_{k-1} + \hat{\mbf{M}}_{k-1} \bm{\delta}_{k-1} + \hat{\mbf{G}}_{k-1} \bm{\xi}_{k-1}$, $\mbf{w}_{k-1} = \mbf{c}_w + \mbf{S}_w \bm{\vartheta}_{k-1} + \mbf{G}_w \bm{\varphi}_{k-1}$, and $\mbf{w}_{k} = \mbf{c}_w + \mbf{S}_w \bm{\vartheta}_k + \mbf{G}_w \bm{\varphi}_{k}$. Moreover, since $\check{\mbf{z}}_k \in \realset^{n - n_z}$, then there must exist $\bm{\delta}_\text{R} \in \realset^{n - n_z}$ such that $\check{\mbf{z}}_{k} = \zeros{(n-n_z)}{1} + \eye{n-n_z} \bm{\delta}_\text{R} \in (\eye{n-n_z}, \noarg, \zeros{(n-n_z)}{1})\lzon$. Then, substituting these equalities in \eqref{eq:desc_systemSVDdynamicslz} leads to
\begin{align}
\begin{bmatrix} \tilde{\mbf{z}}_{k} \\ \check{\mbf{z}}_{k} \end{bmatrix} & = \begin{bmatrix} \tilde{\mbf{A}} \hat{\mbf{c}}_{k-1} + \tilde{\mbf{B}} \mbf{u}_{k-1} + \tilde{\mbf{B}}_w \mbf{c}_{w} \\ \zeros{(n-n_z)}{1} \end{bmatrix} + \begin{bmatrix} \tilde{\mbf{A}} \hat{\mbf{G}}_{k-1} & \tilde{\mbf{B}}_w \mbf{G}_{w} \\ \mbf{0} & \mbf{0} \end{bmatrix} \begin{bmatrix} \bm{\xi}_{k-1} \\ \bm{\varphi}_{k-1} \end{bmatrix} \nonumber \\ & + \begin{bmatrix} \tilde{\mbf{A}} \hat{\mbf{M}}_{k-1} & \tilde{\mbf{B}}_w \mbf{M}_{w} & \bm{0} \\ \mbf{0} & \mbf{0} & \eye{n-n_z} \end{bmatrix} \begin{bmatrix} \bm{\delta}_{k-1} \\ \bm{\vartheta}_{k-1} \\ \bm{\delta}_\text{R} \end{bmatrix}. \label{eq:desc_lzlema1proof1}
\end{align}		
From the constraint \eqref{eq:desc_systemSVDconstraintslz}, we have that
\begin{align}
& \check{\mbf{B}}_w \mbf{c}_w +  \check{\mbf{B}}_w \mbf{G}_w \bm{\varphi}_k + \check{\mbf{B}}_w \mbf{M}_w \bm{\vartheta}_k + \check{\mbf{B}} \mbf{u}_{k} \nonumber \\ & + \check{\mbf{A}} \begin{bmatrix} \tilde{\mbf{A}} \hat{\mbf{c}}_{k-1} + \tilde{\mbf{B}} \mbf{u}_{k-1} + \tilde{\mbf{B}}_w \mbf{c}_{w}  \\  \zeros{(n-n_z)}{1} \end{bmatrix}  + \check{\mbf{A}} \begin{bmatrix} \tilde{\mbf{A}} \hat{\mbf{G}}_{k-1} & \tilde{\mbf{B}}_w \mbf{G}_{w} \\ \mbf{0} & \mbf{0} \end{bmatrix} \begin{bmatrix} \bm{\xi}_{k-1} \\ \bm{\varphi}_{k-1} \end{bmatrix} \nonumber \\
& + \check{\mbf{A}} \begin{bmatrix} \tilde{\mbf{A}} \hat{\mbf{M}}_{k-1} & \tilde{\mbf{B}}_w \mbf{M}_{w} & \bm{0} \\ \mbf{0} & \mbf{0} & \eye{n-n_z} \end{bmatrix} \begin{bmatrix} \bm{\delta}_{k-1} \\ \bm{\vartheta}_{k-1} \\ \bm{\delta}_\text{R} \end{bmatrix} = \mbf{0}. \label{eq:desc_lzlema1proof2}
\end{align}		
Rearranging the equalities \eqref{eq:desc_lzlema1proof1} and \eqref{eq:desc_lzlema1proof2}, grouping the variables $(\bm{\delta}_{k-1}, \bm{\vartheta}_{k-1}, \bm{\vartheta}_k, \bm{\delta}_\text{R})$ and $(\bm{\xi}_{k-1}, \bm{\varphi}_{k-1}, \bm{\varphi}_{k})$, and writing the result in the CLG-rep \eqref{eq:desc_tgrep}, proves the lemma.\qed

\begin{remark} \rm The enclosure $\bar{Z}_k$ given by Lemma \ref{lem:desc_predictionlz} has $n_\delta + n_{\delta_\text{R}} + 2 n_{\delta_w}$ lines, $n_g + 2 n_{g_w}$ generators, and $n_c + 2 n_{c_w} + n - n_z$ constraints.
\end{remark}

Lemma \ref{lem:desc_predictionlz} provides a predicted enclosure for the state $\mbf{z}_k$ in which the equality constraints \eqref{eq:desc_systemSVDconstraintslz} are directly incorporated, thus satisfying \eqref{eq:predictionlds}. Similarly to CZs, this is made possible by the fact that LZs incorporate equality constraints. The prediction-update algorithm proposed for \eqref{eq:desc_system} consists in the computation, at each $k \geq 1$, of LZs $\bar{Z}_k$, $\hat{Z}_k$, and $\hat{X}_k$, such that $\bar{Z}_k = (\bar{\mbf{M}}_k, \bar{\mbf{G}}_k, \bar{\mbf{c}}_k, \bar{\mbf{S}}_k, \bar{\mbf{A}}_k, \bar{\mbf{b}}_k)\lzon$, $\hat{Z}_k = \bar{Z}_k \cap_{\mbf{C}\mbf{T}} ((\mbf{y}_k - \mbf{D}_u \mbf{u}_k) \oplus (-\mbf{D}_v V_k))$, and $\hat{X}_k = \mbf{T} \hat{Z}_k$. The initial set is $\hat{Z}_0$. The algorithm operates recursively with $(\bar{Z}_k,\hat{Z}_k)$ in the state-space \eqref{eq:desc_systemSVDlz}, while the enclosure in the original state-space \eqref{eq:desc_system} is given by $\hat{X}_k$.

\begin{remark} \rm \label{rem:desc_lzestconvervativeness}
	By construction, the CLG-rep in Lemma \ref{lem:desc_predictionlz} represents the exact feasible state set of  \eqref{eq:desc_systemSVDlz} at time $k$, given the known state and uncertainty bounds. Additionally, $\hat{Z}_k$ and $\hat{X}_k$ can be computed exactly. However, to reduce the complexity of the resulting sets in practice, these are outer approximated using the complexity reduction algorithms in Section \ref{sec:desc_lzcomplexityreduction}. As a result, by applying Algorithm \ref{alg:desc_lzugennelim}, the computation of enclosures for descriptor systems with reduced set complexity using LZs is less conservative compared to CZs. This is because the line elimination method leads to an equivalent enclosure, avoiding the introduction of conservatism.
\end{remark}

\begin{remark} \rm The CZ method proposed in \cite{Rego2020b} requires the knowledge of a bounded set enclosing $\mbf{x}_k$. In contrast, state estimation of  \eqref{eq:desc_system} using LZs does not require such a bounded set. As a result, the new method can be applied also to unstable LDS. Moreover, since $X_0$ is described in CLG-rep \eqref{eq:desc_tgrep}, our method does not require prior knowledge of a initial bounded set $X_0$. The enclosures $\hat{X}_k$ will be bounded based on the system's observability. Thanks to their properties, LZs can still provide consistent sets even when full observability is not achieved (sets may be unbounded in some components).
\end{remark}

\section{Open-loop active fault diagnosis of LDS using LZs}  \label{sec:AFDLZ}

The work in \cite{Rego2020b} addressed the problem of set-based AFD for LDS using constrained zonotopes, enabling the separation of the final output reachable set over a given time interval. In this work, we adopt the concept of reachable tubes, which reduces conservatism in input separation and enables more accurate fault diagnosis based on the entire output sequence. We apply the method developed in Section \ref{sec:desc_estimationLZ} to design a tube-based AFD approach that accounts for a finite number of potential abrupt faults, utilizing LZs. This approach allows for the separation of output reachable tubes, including unbounded ones, at any given time instant within the time interval.

Let\footnote{For AFD, we adopt an open-loop approach. To refine the known bound $W$ using the algebraic constraints at time $k$, we augment the state vector with the disturbances $\mbf{w}_k$. While this augmentation could also be incorporated into state estimation, our experiments showed no significant improvement, as the update step with $\mbf{y}_k$ already provides sufficient correction.} $\mbf{z}_k = (\tilde{\mbf{z}}_k,\check{\mbf{z}}_k) = (\inv{\mbf{T}} \mbf{x}_k, \mbf{w}_k)$, $\tilde{\mbf{z}}_k \in \realset^{n_z}$, $\check{\mbf{z}}_k \in \realset^{\check{n}_z}$,
with $\check{n}_z \triangleq n+n_w-n_z$,  $\mbf{T}^{[i]} = \inv{((\mbf{V}^{[i]})^T)}$, $\mbf{V}^{[i]}$ being obtained from the SVD $\mbf{E}^{[i]} = \mbf{U}^{[i]} \mbf{\Sigma}^{[i]} (\mbf{V}^{[i]})^T$. Then, \eqref{eq:desc_system} can be rewritten as
\begin{subequations} \label{eq:desc_systemSVDfaultlz}
	\begin{align} 
	\tilde{\mbf{z}}_k^{[i]} & = \tilde{\mbf{A}}_z^{[i]} \mbf{z}_{k-1}^{[i]} + \tilde{\mbf{B}}^{[i]} \mbf{u}_{k-1}, \label{eq:systemSVDfaultdynamicslz} \\
	\mbf{0} & = \check{\mbf{A}}_z^{[i]} \mbf{z}_{k}^{[i]} + \check{\mbf{B}}^{[i]} \mbf{u}_{k}, \label{eq:desc_systemSVDfaultconstraintslz} \\
	\mbf{y}_k^{[i]} & = \mbf{F}^{[i]} \mbf{z}_k^{[i]} + \mbf{D}^{[i]} \mbf{u}_{k} + \mbf{D}_v^{[i]} \mbf{v}_{k}, \label{eq:systemSVDfaultoutputlz}
	\end{align}
\end{subequations}
with $\mbf{F}^{[i]} = \mbf{C}^{[i]} \mbf{T}^{[i]} \mbf{L}$, where $\mbf{L} = [ \eye{n} \,\; \zeros{n}{n_w}]$, $\tilde{\mbf{A}}_z^{[i]} = [\tilde{\mbf{A}}^{[i]} \,\; \tilde{\mbf{B}}_w^{[i]}]$, and $\check{\mbf{A}}_z^{[i]} = [\check{\mbf{A}}^{[i]} \,\; \check{\mbf{B}}_w^{[i]}]$. 
For each model $i \in \modelset$, variables $\tilde{(\cdot)}$ and $\check{(\cdot)}$  are defined according to \eqref{eq:desc_SVDmatriceslz}. Moreover, let $Z_\sigma^{[i]} = (\mbf{M}_\sigma^{[i]},  \mbf{G}_\sigma^{[i]}, \mbf{c}_\sigma^{[i]}, \mbf{S}_\sigma^{[i]},$ $\mbf{A}_\sigma^{[i]}, \mbf{b}_\sigma^{[i]})\lzon \triangleq \inv{(\mbf{T}^{[i]})} X_0 \times W $, and define the initial feasible set $Z_0^{[i]} (\mbf{u}_0) = \{\mbf{z} \in Z_\sigma^{[i]} : \eqref{eq:desc_systemSVDfaultconstraintslz} \text{ holds for }k=0\}$, which is given by $Z_0^{[i]}(\mbf{u}_0) = (\mbf{M}_0^{[i]}, \mbf{G}_0^{[i]},\mbf{c}_0^{[i]},\mbf{S}_0^{[i]},\mbf{A}_0^{[i]}, \mbf{b}_0^{[i]}(\mbf{u}_0))\lzon$, where $\mbf{M}_0^{[i]} = \mbf{M}_\sigma^{[i]}$, $\mbf{G}_0^{[i]} = \mbf{G}_\sigma^{[i]}$, $\mbf{c}_0^{[i]} = \mbf{c}_\sigma^{[i]}$,
\begin{equation} \label{eq:initialAblz}
\mbf{S}_0^{[i]} = \begin{bmatrix} \mbf{S}_\sigma^{[i]} \\ \check{\mbf{A}}_z^{[i]} \mbf{M}_{0}^{[i]} \end{bmatrix}\! , \, \mbf{A}_0^{[i]} = \begin{bmatrix} \mbf{A}_\sigma^{[i]} \\ \check{\mbf{A}}_z^{[i]} \mbf{G}_{0}^{[i]} \end{bmatrix}\! , \, \mbf{b}_0^{[i]}(\mbf{u}_0) = \begin{bmatrix} \mbf{b}_\sigma^{[i]} \\ -\check{\mbf{A}}_z^{[i]} \mbf{c}_0^{[i]} {-} \check{\mbf{B}}^{[i]} \mbf{u}_{0} \end{bmatrix}.
\end{equation}

Let $\seq{\mbf{u}} = (\mbf{u}_0, ..., \mbf{u}_N)\in \mathbb{R}^{(N+1)n_u}$, $\seq{\mbf{w}} = (\mbf{w}_0, \ldots$, $\mbf{w}_N)\in \mathbb{R}^{(N+1)n_w}$, $\seq{\mbf{v}} = (\mbf{v}_0, \ldots$, $\mbf{v}_N)\in \mathbb{R}^{(N+1)n_v}$, $\seq{W} = W \times \ldots \times W$, and $\seq{V} = V \times \ldots V$. Define the solution mappings $(\bm{\phi}_k^{[i]},\bm{\psi}_k^{[i]}) : \realset^{(k+1)n_u} \times \realset^n \times \realset^{(k+1)n_w} \times \realset^{n_v} \to \realset^{n+n_w} \times \realset^{n_y}$ such that $\bm{\phi}_k^{[i]}(\seq{\mbf{u}},\mbf{z}_0, \seq{\mbf{w}})$ and $\bm{\psi}_k^{[i]}(\seq{\mbf{u}},\mbf{z}_0, \seq{\mbf{w}}, \mbf{v}_k)$ are the state and output of \eqref{eq:desc_systemSVDfaultlz} at time $k$, respectively. Then, for each $i \in \modelset$, define \emph{state and output reachable sets} at time $k$ as $Z_k^{[i]}(\seq{\mbf{u}}) \triangleq \{ \bm{\phi}_k^{[i]}(\seq{\mbf{u}},\mbf{z}_0, \seq{\mbf{w}}) : (\mbf{z}_0, \seq{\mbf{w}}) \in Z_0^{[i]}(\mbf{u}_0) \times \seq{W} \}$, and $Y_k^{[i]}(\seq{\mbf{u}}) \triangleq \{ \bm{\psi}_k^{[i]}(\seq{\mbf{u}},\mbf{z}_0, \seq{\mbf{w}}, \mbf{v}_k) : (\mbf{z}_0,\seq{\mbf{w}},\mbf{v}_k) \in Z_0^{[i]}(\mbf{u}_0) \times \seq{W} \times V\}$, respectively. In addition, let $\seq{\bm{\phi}}\zspace^{[i]}(\seq{\mbf{u}}, \mbf{z}_0, \seq{\mbf{w}}) \triangleq (\bm{\phi}_0^{[i]},\bm{\phi}_1^{[i]},\ldots,\bm{\phi}_N^{[i]})(\seq{\mbf{u}},\mbf{z}_0, \seq{\mbf{w}})$, and $\seq{\bm{\psi}}\zspace^{[i]}(\seq{\mbf{u}},\mbf{z}_0,$ $\seq{\mbf{w}},\seq{\mbf{v}}) \triangleq (\bm{\psi}_0^{[i]},\bm{\psi}_1^{[i]},\ldots,\bm{\psi}_N^{[i]})(\seq{\mbf{u}},\mbf{z}_0, \seq{\mbf{w}}, \seq{\mbf{v}})$, with $N \in \naturalset$. Given $k \in [0,N]$, for each model $i \in \modelset$, we define the \emph{state reachable tube} (SRT) and the \emph{output reachable tube} (ORT) as $\seq{Z}\zspace^{[i]}(\seq{\mbf{u}}) \triangleq \{ \seq{\bm{\phi}}\zspace^{[i]}(\seq{\mbf{u}},\mbf{z}_0, \seq{\mbf{w}}) : (\mbf{z}_0, \seq{\mbf{w}}) \in Z_0^{[i]}(\mbf{u}_0) \times \seq{W} \}$ and $\seq{Y}\zspace^{[i]}(\seq{\mbf{u}}) \triangleq  \{ \seq{\bm{\psi}}\zspace^{[i]}(\seq{\mbf{u}},\mbf{z}_0, \seq{\mbf{w}}, \seq{\mbf{v}}) : (\mbf{z}_0,\seq{\mbf{w}},\seq{\mbf{v}}) \in Z_0^{[i]}(\mbf{u}_0) \times \seq{W} \times \seq{V}\}$, respectively, which are denoted as $\zseq{Z}\lzon^{[i]}(\seq{\mbf{u}})$ and $\zseq{Y}\lzon^{[i]}(\seq{\mbf{u}})$ if obtained using LZs. Moreover, let $\realset^{n-n_z} \triangleq (\eye{n-n_z},\noarg,\zeros{(n-n_z)}{1})\lzon$, $W = (\mbf{M}_w, \mbf{G}_w, \mbf{c}_w, $ $\mbf{S}_w, \mbf{A}_w, \mbf{b}_w)\lzon$, and $Z_\text{A} = (\mbf{M}_\text{A}, \mbf{G}_\text{A}, \mbf{c}_\text{A}, \mbf{S}_\text{A}, \mbf{A}_\text{A}, \mbf{b}_\text{A})\lzon \triangleq \realset^{n-n_z} \times W$, where $\mbf{c}_\text{A} = (\zeros{(n-n_z)}{1}, \mbf{c}_w)$, $\mbf{G}_\text{A} = [\mbf{0} \,\; \mbf{G}_w^T]^T$, $\mbf{M}_\text{A} = \bdiag{\eye{n-n_z}, \mbf{M}_w}$, $\mbf{S}_\text{A} = [\mbf{0} \,\; \mbf{S}_w]$, $\mbf{A}_\text{A} = \mbf{A}_w$, $\mbf{b}_\text{A} = \mbf{b}_w$.  %
Using \eqref{eq:desc_lzlimage}--\eqref{eq:desc_lzmsum}, the set $Z_k^{[i]}(\seq{\mbf{u}})$ is given by the CLG-rep $( \mbf{M}_k^{[i]}, \mbf{G}_k^{[i]}, \mbf{c}_k^{[i]}(\seq{\mbf{u}}), \mbf{S}_k^{[i]}, \mbf{A}_k^{[i]},$ $\mbf{b}_k^{[i]}(\seq{\mbf{u}}))\lzon$, 
where $\mbf{M}_k^{[i]}$, $\mbf{G}_k^{[i]}$, $\mbf{c}_k^{[i]}(\seq{\mbf{u}})$, $\mbf{S}_k^{[i]}$, $\mbf{A}_k^{[i]}$, and $\mbf{b}_k^{[i]}(\seq{\mbf{u}})$ are given by 
\begin{align*}
& \mbf{c}_k^{[i]}(\seq{\mbf{u}}) = \begin{bmatrix} \tilde{\mbf{A}}_z^{[i]} \mbf{c}_{k-1}^{[i]}(\seq{\mbf{u}}) + \tilde{\mbf{B}}^{[i]} \mbf{u}_{k-1} \\ \mbf{c}_\text{A} \end{bmatrix}, \; \mbf{S}_k^{[i]} = \begin{bmatrix} \bdiag{\mbf{S}_{k-1}^{[i]},\mbf{S}_\text{A}} \\ \check{\mbf{A}}_z^{[i]} \mbf{M}_{k}^{[i]} \end{bmatrix}, \\
& \mbf{G}_k^{[i]} = \bdiag{\tilde{\mbf{A}}_z^{[i]} \mbf{G}_{k-1}^{[i]},\mbf{G}_\text{A}}, \; \mbf{M}_k^{[i]} = \bdiag{\tilde{\mbf{A}}_z^{[i]} \mbf{M}_{k-1}^{[i]},\mbf{M}_\text{A}}, \\
& \mbf{A}_k^{[i]} = \begin{bmatrix} \bdiag{\mbf{A}_{k-1}^{[i]},\mbf{A}_\text{A}} \\ \check{\mbf{A}}_z^{[i]} \mbf{G}_{k}^{[i]} \end{bmatrix}, \; \mbf{b}_k^{[i]}(\seq{\mbf{u}}) = \begin{bmatrix} \verti{\mbf{b}_{k-1}^{[i]}(\seq{\mbf{u}}),\mbf{b}_\text{A}} \\ -\check{\mbf{A}}_z^{[i]} \mbf{c}_k^{[i]}(\seq{\mbf{u}}) - \check{\mbf{B}}^{[i]} \mbf{u}_{k} \end{bmatrix}, 
\end{align*}
for $k = 1,2,\ldots,N$. It is worth noting that the third constraint in $(\mbf{S}_k^{[i]}, \mbf{A}_k^{[i]}, \mbf{b}_k^{[i]}(\seq{\mbf{u}}))$ comes from the fact that \eqref{eq:desc_systemSVDfaultconstraintslz} must hold. Using the above recursive equations with the initial values \eqref{eq:initialAblz}, reorganizing equations, and removing redundant constraints, the SRT $\zseq{Z}\lzon^{[i]}(\seq{\mbf{u}})$ can be written as an explicit function of the input sequence $\seq{\mbf{u}}$ as $\zseq{Z}\lzon^{[i]}(\seq{\mbf{u}}) = (\seq{\mbf{M}}\zerospace^{[i]}, \zseq{\mbf{G}}^{[i]}, \zseq{\mbf{c}}^{[i]}(\seq{\mbf{u}}), \zseq{\mbf{S}}^{[i]}, \zseq{\mbf{A}}^{[i]}, \zseq{\mbf{b}}^{[i]}(\seq{\mbf{u}}) )\lzon$, where $\zseq{\mbf{c}}^{[i]}(\seq{\mbf{u}}) = \zseq{\mbf{Q}}^{[i]} \mbf{c}_\sigma + \zseq{\mbf{p}}^{[i]} + \zseq{\mbf{H}}^{[i]} \seq{\mbf{u}}$, and $\zseq{\mbf{b}}^{[i]}(\seq{\mbf{u}}) = \zseq{\bm{\alpha}}^{[i]} + \zseq{\bm{\Lambda}}^{[i]} \mbf{c}_\sigma + \zseq{\bm{\Omega}}^{[i]} \seq{\mbf{u}}$ (definitions can be found in \ref{sec:app_statetube}). Since $\zseq{Z}\lzon^{[i]}(\seq{\mbf{u}})$ is a line zonotope, the ORT $\zseq{Y}^{[i]}\lzon(\seq{\mbf{u}})$ is then a LZ obtained in accordance with \eqref{eq:systemSVDfaultoutputlz}, as 
\begin{equation} \label{eq:desc_outputreachablelz}
\zseq{Y}^{[i]}\lzon(\seq{\mbf{u}}) = \seq{\mbf{F}}\zerospace^{[i]} \zseq{Z}\lzon^{[i]}(\seq{\mbf{u}}) \oplus \seq{\mbf{D}}\zerospace^{[i]} \seq{\mbf{u}} \oplus \seq{\mbf{D}}\zerospace^{[i]}_v \seq{V},
\end{equation}
where $\zseq{\mbf{F}}^{[i]} \triangleq 
\bdiag{\mbf{F}^{[i]}, \ldots, \mbf{F}^{[i]}}$, $\seq{\mbf{D}}\zerospace^{[i]} \triangleq 
\bdiag{\mbf{D}^{[i]}, \ldots, \mbf{D}^{[i]}}$, and $\seq{\mbf{D}}\zerospace^{[i]}_v \triangleq 
\bdiag{\mbf{D}^{[i]}_v, \ldots, \mbf{D}^{[i]}_v}$. Using properties \eqref{eq:desc_lzlimage} and \eqref{eq:desc_lzmsum}, and letting $V \triangleq ( \mbf{M}_v, \mbf{G}_v, \mbf{c}_v, \mbf{S}_v, \mbf{A}_v, \mbf{b}_v)\lzon$, the ORT \eqref{eq:desc_outputreachablelz} is given by the CLG-rep $\zseq{Y}\lzon^{[i]}(\seq{\mbf{u}}) = (\zseq{\mbf{M}}^{Y[i]}, \zseq{\mbf{G}}^{Y[i]}, $ $\seq{\mbf{c}}\zspace^{Y[i]}(\seq{\mbf{u}}), \zseq{\mbf{S}}^{Y[i]}, \zseq{\mbf{A}}^{Y[i]}, \zseq{\mbf{b}}^{Y[i]}(\seq{\mbf{u}}) )\lzon$, in which
\begin{align}
& \zseq{\mbf{c}}^{Y[i]}(\seq{\mbf{u}}) = \zseq{\mbf{F}}^{[i]} \zseq{\mbf{c}}^{[i]}(\seq{\mbf{u}}) + \zseq{\mbf{D}}^{[i]} \seq{\mbf{u}} + \zseq{\mbf{D}}^{[i]}_v \seq{\mbf{c}}_v, \label{eq:desc_outputcklz} \\
& \zseq{\mbf{M}}^{Y[i]} = \big[\zseq{\mbf{F}}^{[i]} \zseq{\mbf{M}}^{[i]} \,\; \zseq{\mbf{D}}^{[i]}_v \zseq{\mbf{M}}^{[i]}_v \big],\quad \zseq{\mbf{G}}^{Y[i]} = \big[\zseq{\mbf{F}}^{[i]} \zseq{\mbf{G}}^{[i]} \,\; \zseq{\mbf{D}}^{[i]}_v \zseq{\mbf{G}}^{[i]}_v \big],\\
& \zseq{\mbf{S}}^{Y[i]} = \bdiag{\zseq{\mbf{S}}^{[i]}, \zseq{\mbf{S}}^{[i]}_v},\quad \zseq{\mbf{A}}^{Y[i]} = \bdiag{\zseq{\mbf{A}}^{[i]}, \zseq{\mbf{A}}^{[i]}_v},\\ & \zseq{\mbf{b}}^{Y[i]}(\seq{\mbf{u}}) = \verti{\zseq{\mbf{b}}^{[i]}(\seq{\mbf{u}}),\zseq{\mbf{b}}^{[i]}_v},  \label{eq:desc_outputbklz}
\end{align}
where $\seq{\mbf{c}}_v \triangleq \verti{\mbf{c}_v, \ldots, \mbf{c}_v}$,  $\seq{\mbf{G}}_v \triangleq \bdiag{\mbf{G}_v, \ldots, \mbf{G}_v}$,  $\seq{\mbf{A}}_v \triangleq \bdiag{\mbf{A}_v^{[i]}, \ldots, \mbf{A}_v^{[i]}}$, and $\seq{\mbf{b}}_v \triangleq \verti{\mbf{b}_v, \ldots, \mbf{b}_v}$.

\subsection{Separation of output reachable tubes}

Consider the input sequence $\seq{\mbf{u}} \in \seq{U}$ to be injected into the set of models \eqref{eq:desc_systemSVDfaultlz}, and let $\seq{\mbf{y}}\zspace^{[i]}$ denote the observed output sequence. The following definition is essential for the AFD method proposed in this paper.
\begin{definition} \rm \label{def:separatinginputcz}
	An input sequence $\seq{\mbf{u}}$ is said to be a feasible \emph{separating input} of the tuple $(\seq{Y}\zspace^{[\cdot]}(\seq{\mbf{u}}),\modelset, \seq{U})$ if $\seq{\mbf{u}} \in \seq{U}$, and for every $i,j \in \modelset$, $i \neq j$, $\seq{Y}\zspace^{[i]}(\seq{\mbf{u}}) \cap \seq{Y}\zspace^{[j]}(\seq{\mbf{u}}) = \emptyset$.
\end{definition}
Clearly, if $\seq{Y}\zspace^{[i]}(\seq{\mbf{u}}) \cap \seq{Y}\zspace^{[j]}(\seq{\mbf{u}}) = \emptyset$ holds for all $i,j \in \modelset$, $i \neq j$, then $\seq{\mbf{y}}\zspace^{[i]} \in \zseq{Y}^{[i]}(\seq{\mbf{u}})$ must hold only for one $i$. In the case this is not valid for any $i \in \modelset$, one concludes that the plant dynamics do not belong to the set of models \eqref{eq:desc_systemSVDfaultlz}. We denote the set of inputs satisfying Definition \ref{def:separatinginputcz} as $\mathcal{S}(\seq{Y}\zspace^{[\cdot]}(\seq{\mbf{u}}),\modelset,\seq{U})$. By construction, $\mathcal{S}(\seq{Y}\zspace^{[\cdot]}(\seq{\mbf{u}}),\modelset,\seq{U}) \subseteq \seq{U}$. %
In this section, we are interested in the design of an optimal input sequence that minimizes a functional $J(\seq{\mbf{u}})$  satisfying $\seq{\mbf{u}} \in \mathcal{S}(\zseq{Y}\lzon^{[\cdot]}(\seq{\mbf{u}}),\modelset,\seq{U})$. The following theorem is based on the computation of $\zseq{Y}\lzon^{[i]}(\seq{\mbf{u}})$ by \eqref{eq:desc_outputreachablelz}.

\begin{theorem} \rm \label{thm:desc_separatinginputlz}
	An input $\seq{\mbf{u}} \in \seq{U}$ belongs to $\mathcal{S}(\zseq{Y}\lzon^{[\cdot]}(\seq{\mbf{u}}),\modelset,\seq{U})$ iff
	\begin{equation} \begin{aligned} \label{eq:desc_separationconditionlzline}
	 \begin{bmatrix} \zseq{\mbf{N}}^{(q)} \\ \zseq{\bm{\Omega}}^{(q)} \end{bmatrix} \seq{\mbf{u}} \notin \zseq{\mathcal{Y}}^{(q)} \triangleq \left( \begin{bmatrix} \zseq{\mbf{M}}^{Y(q)} \\ \zseq{\mbf{S}}^{Y(q)} \end{bmatrix}, \begin{bmatrix} \zseq{\mbf{G}}^{Y(q)} \\ \zseq{\mbf{A}}^{Y(q)} \end{bmatrix}, \; \begin{bmatrix} \zseq{\mbf{c}}^{Y(q)} \\ -\zseq{\mbf{b}}^{Y(q)} \end{bmatrix} \right)\lzon \end{aligned}
	\end{equation}
	$\forall q \in \mathcal{Q}$, where $\mathcal{Q}$ is the set of all possible combinations of $(i,j) \in \modelset \times \modelset$, $i < j$, $\zseq{\mbf{N}}^{(q)} \triangleq (\zseq{\mbf{F}}^{[j]} \zseq{\mbf{H}}^{[j]} + \zseq{\mbf{D}}^{[j]}) - (\zseq{\mbf{F}}^{[i]} \zseq{\mbf{H}}^{[i]} + \zseq{\mbf{D}}^{[i]})$, $\zseq{\bm{\Omega}}^{(q)} \triangleq [(\zseq{\bm{\Omega}}^{[i]})^T \; \mbf{0} \; (\zseq{\bm{\Omega}}^{[j]})^T \; \mbf{0}]^T$, and $\zseq{\mbf{M}}^{Y(q)} \triangleq [\zseq{\mbf{M}}^{Y[i]} \,\; - \zseq{\mbf{M}}^{Y[j]}]$, $\zseq{\mbf{G}}^{Y(q)} \triangleq [\zseq{\mbf{G}}^{Y[i]} \,\; - \zseq{\mbf{G}}^{Y[j]}]$, $\zseq{\mbf{S}}^{Y(q)} \triangleq \bdiag{\zseq{\mbf{S}}^{Y[i]}, \zseq{\mbf{S}}^{Y[j]}}$, $\zseq{\mbf{A}}^{Y(q)} \triangleq \bdiag{\zseq{\mbf{A}}^{Y[i]}, \zseq{\mbf{A}}^{Y[j]}}$, $\zseq{\mbf{c}}^{Y(q)} \triangleq \zseq{\mbf{c}}^{Y[i]}(\seq{\mbf{0}}) - \zseq{\mbf{c}}^{Y[j]}(\seq{\mbf{0}})$, $\zseq{\mbf{b}}^{Y(q)} \triangleq \verti{\zseq{\mbf{b}}^{Y[i]}(\seq{\mbf{0}}),\zseq{\mbf{b}}^{Y[j]}(\seq{\mbf{0}})}$, with $i,j \in \modelset$ being the pair associated with each $q \in \mathcal{Q}$, and $\seq{\mbf{0}}$ denotes the zero input sequence.
\end{theorem}

\proof
From \eqref{eq:desc_outputcklz} and \eqref{eq:desc_outputbklz}, and the expressions obtained for $\zseq{\mbf{c}}^{[i]}(\seq{\mbf{u}})$ and $\zseq{\mbf{b}}^{[i]}(\seq{\mbf{u}})$, one has $\zseq{\mbf{c}}^{Y[i]}(\mbf{0}) = \zseq{\mbf{F}}^{[i]} (\zseq{\mbf{Q}}^{[i]} \mbf{c}_z^{[i]} + \zseq{\mbf{p}}^{[i]}) + \zseq{\mbf{D}}^{[i]}_v \seq{\mbf{c}}_v$, and $\zseq{\mbf{b}}^{Y[i]}(\seq{\mbf{0}}) = [(\zseq{\bm{\alpha}}^{[i]} + \zseq{\bm{\Lambda}}^{[i]} \mbf{c}_z)^T \,\; (\seq{\mbf{b}}_v)^T]^T$. Hence, the following equalities hold: %
\begin{align}
& \seq{\mbf{c}}\zerospace^{Y[i]}(\seq{\mbf{u}}) = \seq{\mbf{c}}\zerospace^{Y[i]}(\mbf{0}) + (\seq{\mbf{F}}\zerospace^{[i]} \seq{\mbf{H}}\zerospace^{[i]} + \seq{\mbf{D}}\zerospace^{[i]}) \seq{\mbf{u}}, \label{eq:desc_outputcseq0} \\
& \seq{\mbf{b}}\zerospace^{Y[i]}(\seq{\mbf{u}}) = \seq{\mbf{b}}\zerospace^{Y[i]}(\mbf{0}) + \begin{bmatrix} \seq{\bm{\Omega}}\zerospace^{[i]} \\ \mbf{0} \end{bmatrix} \seq{\mbf{u}}. \label{eq:desc_outputbseq0}
\end{align}
From \eqref{eq:desc_lzintersection} with $\mbf{R} = \mbf{I}$, $\zseq{Y}\lzon^{[i]}(\seq{\mbf{u}}) \cap \zseq{Y}\lzon^{[j]}(\seq{\mbf{u}}) = \emptyset$ iff $\nexists (\bm{\delta}^{(q)}, \bm{\xi}^{(q)}) \in \realset \times B_\infty$ such that
\begin{equation*}
\begin{bmatrix} \zseq{\mbf{M}}^{Y[i]} & -\seq{\mbf{M}}\zerospace^{Y[j]} \\ \zseq{\mbf{S}}^{Y[i]} & \mbf{0} \\ \mbf{0} & \zseq{\mbf{S}}^{Y[j]} \end{bmatrix} \bm{\delta}^{(q)} + \begin{bmatrix} \zseq{\mbf{G}}^{Y[i]} & -\zseq{\mbf{G}}^{Y[j]}  \\ \zseq{\mbf{A}}^{Y[i]} & \mbf{0} \\\mbf{0} & \zseq{\mbf{A}}^{Y[j]} \end{bmatrix} \bm{\xi}^{(q)} = \begin{bmatrix} \zseq{\mbf{c}}^{Y[j]}(\seq{\mbf{u}}) - \zseq{\mbf{c}}^{Y[i]}(\seq{\mbf{u}}) \\ \zseq{\mbf{b}}^{Y[i]}(\seq{\mbf{u}}) \\ \zseq{\mbf{b}}^{Y[j]}(\seq{\mbf{u}}) \end{bmatrix}.
\end{equation*}
Using \eqref{eq:desc_outputcseq0}--\eqref{eq:desc_outputbseq0}, one has $\zseq{\mbf{c}}^{Y[j]}(\seq{\mbf{u}}) - \zseq{\mbf{c}}^{Y[i]}(\seq{\mbf{u}}) = - \zseq{\mbf{c}}^{Y(q)} + \zseq{\mbf{N}}^{(q)} \seq{\mbf{u}}$, and $[(\zseq{\mbf{b}}^{Y[i]}(\seq{\mbf{u}}))^T  \;  (\zseq{\mbf{b}}^{Y[j]}(\seq{\mbf{u}}))^T]^T $ $ \!=\! \zseq{\mbf{b}}^{Y(q)}$ $+ \zseq{\bm{\Omega}}^{(q)} \seq{\mbf{u}}$,
with $\zseq{\mbf{c}}^{Y(q)}$, $\zseq{\mbf{b}}^{Y(q)}$, $\zseq{\mbf{N}}^{(q)}$, and $\zseq{\bm{\Omega}}^{(q)}$ defined as in the statement of the theorem. Then, $\zseq{Y}\lzon^{[i]}(\seq{\mbf{u}}) \cap \zseq{Y}\lzon^{[j]}(\seq{\mbf{u}}) = \emptyset$ iff $\nexists (\bm{\delta}^{(q)},\bm{\xi}^{(q)}) \in \realset \times B_\infty$ such that $\zseq{\mbf{M}}^{Y(q)} \bm{\delta} + \zseq{\mbf{G}}^{Y(q)} \bm{\xi}^{(q)} = - \zseq{\mbf{c}}^{Y(q)} + \zseq{\mbf{N}}^{(q)} \seq{\mbf{u}}$, and $\zseq{\mbf{S}}^{Y(q)} \bm{\delta} + \zseq{\mbf{A}}^{Y(q)} \bm{\xi}^{(q)} = \zseq{\mbf{b}}^{Y(q)} + \zseq{\bm{\Omega}}^{(q)} \seq{\mbf{u}}$, with $\zseq{\mbf{M}}^{Y(q)}$, $\zseq{\mbf{G}}^{Y(q)}$, $\zseq{\mbf{S}}^{Y(q)}$, and $\zseq{\mbf{A}}^{Y(q)}$ defined as in the statement of the theorem. This is true iff $\nexists (\bm{\delta}^{(q)},\bm{\xi}^{(q)}) \in \realset \times B_\infty$ such that
\begin{equation*}
\begin{bmatrix} \zseq{\mbf{M}}^{Y(q)} \\ \zseq{\mbf{S}}^{Y(q)} \end{bmatrix} \bm{\delta}^{(q)} + \begin{bmatrix} \zseq{\mbf{G}}^{Y(q)} \\ \zseq{\mbf{A}}^{Y(q)} \end{bmatrix} \bm{\xi}^{(q)} + \begin{bmatrix} \zseq{\mbf{c}}^{Y(q)} \\ - \zseq{\mbf{b}}^{Y(q)} \end{bmatrix} = \begin{bmatrix} \zseq{\mbf{N}}^{(q)} \\ \zseq{\bm{\Omega}}^{(q)} \end{bmatrix} \seq{\mbf{u}}, 
\end{equation*}
which in turn holds iff \eqref{eq:desc_separationconditionlzline} is satisfied.\qed

Since $\zseq{\mathcal{Y}}^{(q)}$ is a line zonotope, then for a given $q \in \mathcal{Q}$, the relation \eqref{eq:desc_separationconditionlzline} can be verified by solving a linear program (LP).

\begin{lemma} \rm \label{thm:separationlzLP} Let $\mbf{u} \in \seq{U}$, and consider $\zseq{\mbf{N}}^{(q)}$, $\zseq{\bm{\Omega}}^{(q)}$, $\zseq{\mbf{M}}^{Y(q)}$, $\zseq{\mbf{S}}^{Y(q)}$, $\zseq{\mbf{G}}^{Y(q)}$, $\zseq{\mbf{A}}^{Y(q)}$, $\zseq{\mbf{c}}^{Y(q)}$, and $\zseq{\mbf{b}}^{Y(q)}$ as defined in Theorem \ref{thm:desc_separatinginputlz}. Define
    \begin{equation} \label{eq:separationconditionlzLP}
    \hat{\kappa}^{(q)}(\seq{\mbf{u}}) \triangleq \left\{ \begin{aligned} & \underset{\kappa^{(q)}, \bm{\delta}^{(q)}, \bm{\xi}^{(q)}}{\min} \kappa^{(q)} \quad \text{ s.t. } \quad \ninf{\bm{\xi}^{(q)}} \leq 1 + \kappa^{(q)},\\ & 	\begin{bmatrix} \zseq{\mbf{N}}^{(q)} \\ \zseq{\bm{\Omega}}^{(q)} \end{bmatrix} \seq{\mbf{u}} =  \begin{bmatrix} \zseq{\mbf{M}}^{Y(q)} \\ \zseq{\mbf{S}}^{Y(q)} \end{bmatrix} \bm{\delta}^{(q)} + \begin{bmatrix} \zseq{\mbf{G}}^{Y(q)} \\ \zseq{\mbf{A}}^{Y(q)} \end{bmatrix} \bm{\xi}^{(q)} + \begin{bmatrix} \zseq{\mbf{c}}^{Y(q)} \\ -\zseq{\mbf{b}}^{Y(q)} \end{bmatrix} \end{aligned} \right\}.
\end{equation}
Then, \eqref{eq:desc_separationconditionlzline} holds iff $\hat{\kappa}^{(q)}(\seq{\mbf{u}}) > 0$. 
\end{lemma}
\proof
The proof is analogous to Lemma 4 in \cite{Scott2014}. The addition of the unbounded decision variables $\bm{\delta}^{(q)}$ does not result in new facts to the analysis.
\qed

\begin{remark} \rm \label{rem:inputconstraintelimlz}
    The number of decision variables in Lemma \ref{thm:separationlzLP} increases with the dimensionality of the LZ $\zseq{\mathcal{Y}}^{(q)}$ in \eqref{eq:desc_separationconditionlzline}. However, in contrast to the CZ method in \cite{Raimondo2016}, the reduction of the number of constraints and decision variables in \eqref{eq:separationconditionlzLP} is not trivial since $\zseq{\bm{\Omega}}^{(q)} \neq \mbf{0}$ in general.
\end{remark}

\subsection{Input-dependent complexity reduction}\label{sec:inputconstraintelimlz}

To mitigate the complexity of \eqref{eq:separationconditionlzLP}, the number of generators of $\zseq{\mathcal{Y}}^{(q)}$ in \eqref{eq:desc_separationconditionlzline} can be reduced by applying the generator reduction methods discussed in Section \ref{sec:desc_lzcomplexityreduction}. However, this process is limited by the dimension of $\zseq{\mathcal{Y}}^{(q)}$. Therefore, to overcome this limitation, we propose a methodology to reduce the number of equality constraints in the LP \eqref{eq:separationconditionlzLP}. It can be demonstrated that
\begin{equation} \label{eq:faultdiagLZ_celim1}
\begin{aligned}
	& \begin{bmatrix} \zseq{\mbf{N}}^{(q)} \\ \zseq{\bm{\Omega}}^{(q)} \end{bmatrix} \seq{\mbf{u}} \notin \left( \begin{bmatrix} \zseq{\mbf{M}}^{Y(q)} \\ \zseq{\mbf{S}}^{Y(q)} \end{bmatrix}, \begin{bmatrix} \zseq{\mbf{G}}^{Y(q)} \\ \zseq{\mbf{A}}^{Y(q)} \end{bmatrix}, \; \begin{bmatrix} \zseq{\mbf{c}}^{Y(q)} \\ -\zseq{\mbf{b}}^{Y(q)} \end{bmatrix} \right)\lzon \iff \\ & \mbf{0} \notin (\zseq{\mbf{M}}^{Y(q)}, \zseq{\mbf{G}}^{Y(q)}, \zseq{\mbf{c}}^{Y(q)} - \zseq{\mbf{N}}^{(q)} \seq{\mbf{u}}, \zseq{\mbf{S}}^{Y(q)}, \zseq{\mbf{A}}^{Y(q)}, \zseq{\mbf{b}}^{Y(q)} + \zseq{\bm{\Omega}}^{(q)} \seq{\mbf{u}} )\lzon.
\end{aligned}
\end{equation}
Note that the right-hand-side of \eqref{eq:faultdiagLZ_celim1} is a line zonotope, and therefore the number of generators can be reduced as described in Section \ref{sec:desc_lzcomplexityreduction}. However, the constraints in this line zonotope are dependent on the input sequence $\seq{\mbf{u}}$, which is yet to be designed. To address this, we rewrite the right-hand-side of \eqref{eq:faultdiagLZ_celim1} using a change of variables as follows.

By assumption, $\seq{U}$ is a bounded convex polytopic set, thus let $\seq{U} \triangleq ( \seq{\mbf{G}}_u, \seq{\mbf{c}}_u, \seq{\mbf{A}}_u, \seq{\mbf{b}}_u )\czon$. Then, $\seq{\mbf{u}} \in \mathcal{S}(\zseq{Y}\lzon^{[\cdot]}(\seq{\mbf{u}}),\modelset,$ $\seq{U}) \implies \seq{\mbf{u}} \in \seq{U}$ $\iff$ $\exists \seq{\bm{\xi}}_u \in B_\infty(\seq{\mbf{A}}_u, \seq{\mbf{b}}_u)$ such that $\seq{\mbf{u}} = \seq{\mbf{c}}_u +  \seq{\mbf{G}}_u \seq{\bm{\xi}}_u$. Let the right-hand-side in \eqref{eq:faultdiagLZ_celim1} have $n_\delta^Y$ lines and $n_g^Y$ generators, and let $\seq{n}_{g_u}$ be the number of generators of $\seq{U}$. Hence, for such $\seq{\bm{\xi}}_u$,  \eqref{eq:faultdiagLZ_celim1} $\iff \nexists (\bm{\delta}^Y, \bm{\xi}^Y) \in \realset^{n_{\delta}^Y} \times B_\infty^{n_g^Y}$ such that
\begin{align} 
& \left\{ \begin{aligned} 
\mbf{0} & = \zseq{\mbf{M}}^{Y(q)} \bm{\delta}^Y + \zseq{\mbf{G}}^{Y(q)} \bm{\xi}^Y + \zseq{\mbf{c}}^{Y(q)} - \zseq{\mbf{N}}^{(q)} (\seq{\mbf{c}}_u +  \seq{\mbf{G}}_u \seq{\bm{\xi}}_u), \\
\mbf{0} & = \zseq{\mbf{S}}^{Y(q)} \bm{\delta}^Y + \zseq{\mbf{A}}^{Y(q)} \bm{\xi}^Y - \zseq{\mbf{b}}^{Y(q)} - \zseq{\bm{\Omega}}^{(q)} (\seq{\mbf{c}}_u +  \seq{\mbf{G}}_u \seq{\bm{\xi}}_u),  \\ \mbf{0} & = \seq{\mbf{A}}_u \seq{\bm{\xi}}_u - \seq{\mbf{b}}_u,
\end{aligned} \right. \nonumber \\
& \begin{aligned}
\iff & \mbf{0} \notin \Big( \zseq{\mbf{M}}^{Y(q)}, [\zseq{\mbf{G}}^{Y(q)} \,\; - \zseq{\mbf{N}}^{(q)} \seq{\mbf{G}}_u ], \zseq{\mbf{c}}^{Y(q)} - \zseq{\mbf{N}}^{(q)} \seq{\mbf{c}}_u,  \\ & \left. \begin{bmatrix} \zseq{\mbf{S}}^{Y(q)} \\ \mbf{0} \end{bmatrix}, \begin{bmatrix} \zseq{\mbf{A}}^{Y(q)} & -\zseq{\bm{\Omega}}^{(q)} \seq{\mbf{G}}_u \\ \mbf{0} & \seq{\mbf{A}}_u \end{bmatrix}, \begin{bmatrix} \zseq{\mbf{b}}^{Y(q)} + \zseq{\bm{\Omega}}^{(q)} \seq{\mbf{c}}_u \\ \seq{\mbf{b}}_u \end{bmatrix} \right)\lzon \!\!\!\!\!.
\end{aligned} \label{eq:faultdiagLZ_celim2}
\end{align} 
Therefore $\seq{\mbf{u}} \in \mathcal{S}(\zseq{Y}\lzon^{[\cdot]}(\seq{\mbf{u}}),\modelset,\seq{U})\iff $\eqref{eq:faultdiagLZ_celim2} for all $q \in \mathcal{Q}$. To simplify the right-hand side of \eqref{eq:faultdiagLZ_celim2}, the reduction methods outlined in Section \ref{sec:desc_lzcomplexityreduction} can be applied.

\begin{remark} \rm
Generators associated with the sequence $\seq{\mbf{u}} = \mbf{c}_u + \mbf{G}_u \seq{\bm{\xi}}_u$ 
must not be removed, only those from $\bm{\delta}^Y$ and $\bm{\xi}^Y$ are eliminated during the dualization process. Otherwise, there may be no $\seq{\bm{\xi}}_u \in B_\infty(\seq{\mbf{A}}_u, \seq{\mbf{b}}_u)$ satisfying $\seq{\mbf{u}} = \seq{\mbf{c}}_u +  \seq{\mbf{G}}_u \seq{\bm{\xi}}_u$, failing to ensure $\seq{\mbf{u}} \in \seq{U}$ and placing $\seq{\mbf{u}}$ in a conservative enclosure of $\seq{U}$.
\end{remark}
Let $\check{Z}^{(q)}$ denote the LZ obtained after applying the complexity reduction methods in Section \ref{sec:desc_lzcomplexityreduction} to the right-hand-side in \eqref{eq:faultdiagLZ_celim2}, with CLG-rep
\begin{equation} \label{eq:constrainteliminationlzreduced}
( \check{\mbf{M}}^{Z(q)}, [\check{\mbf{G}}^{Z(q)} \,\; \check{\mbf{G}}_u^{(q)} ], \check{\mbf{c}}^{Z(q)}, \check{\mbf{S}}^{Z(q)},  [\check{\mbf{A}}^{Z(q)} \,\; \check{\mbf{A}}_u^{(q)}], \check{\mbf{b}}^{Z(q)} )\lzon.
\end{equation}
Note that, since $\check{\mbf{S}}^{Z(q)}=\mbf{0}$\footnote{After line elimination, constraints involve generators only, see Sec. \ref{sec_312}.}, it is always possible to reduce matrix $\check{\mbf{M}}^{Z(q)}$ to a full column-rank matrix\footnote{\label{QRdec} Use, e.g., QR decomposition. Since lines are unbounded, retaining matrix $\mbf{Q}$ from the decomposition is sufficient, as discarding $\mbf{R}$ does not result in information loss.}. $\check{Z}^{(q)}$ contains the right-hand-side of \eqref{eq:faultdiagLZ_celim2} by construction. Consequently, $\mbf{0} \notin \check{Z}^{(q)} \implies $\eqref{eq:faultdiagLZ_celim2}, hence $\mbf{0} \notin \check{Z}^{(q)} ~ \forall q \in \mathcal{Q} \implies \seq{\mbf{u}} \in \mathcal{S}(\zseq{Y}\lzon^{[\cdot]}(\seq{\mbf{u}}),\modelset,\seq{U})$. Finally, $\mbf{0} \notin \check{Z}^{(q)}$ iff, given $\seq{\bm{\xi}}_u$, $\nexists (\bm{\delta}^Z, \bm{\xi}^Z) \in \realset^{n_\delta^Z} \times B_\infty^{n_g^Z}$ so that $\mbf{0} = \check{\mbf{M}}^{Z(q)} \bm{\delta}^Z + \check{\mbf{G}}^{Z(q)} \bm{\xi}^Z + \check{\mbf{G}}_u^{(q)} \seq{\bm{\xi}}_u + \check{\mbf{c}}^{Z(q)}$, and $\mbf{0} =  \check{\mbf{A}}^{Z(q)} \bm{\xi}^Z + \check{\mbf{A}}_u^{(q)} \seq{\bm{\xi}}_u - \check{\mbf{b}}^{Z(q)}$, which holds iff $\nexists (\bm{\delta}^Z, \bm{\xi}^Z) \in \realset^{n_\delta^Z} \times B_\infty^{n_g^Z}$ such that
\begin{align}
&
- \begin{bmatrix} \check{\mbf{G}}_u^{(q)} \\ \check{\mbf{A}}_u^{(q)} \end{bmatrix} \seq{\bm{\xi}}_u = \begin{bmatrix} \check{\mbf{M}}^{Z(q)} \\ \mbf{0} \end{bmatrix} \bm{\delta}^Z + \begin{bmatrix} \check{\mbf{G}}^{Z(q)} \\ \check{\mbf{A}}^{Z(q)} \end{bmatrix} \bm{\xi}^Z + \begin{bmatrix} \check{\mbf{c}}^{Z(q)} \\ - \check{\mbf{b}}^{Z(q)} \end{bmatrix} \nonumber \\
& \iff  - \begin{bmatrix} \check{\mbf{G}}_u^{(q)} \\ \check{\mbf{A}}_u^{(q)} \end{bmatrix} \seq{\bm{\xi}}_u \notin \left(\begin{bmatrix} \check{\mbf{M}}^{Z(q)} \\ \mbf{0} \end{bmatrix}, \begin{bmatrix} \check{\mbf{G}}^{Z(q)} \\ \check{\mbf{A}}^{Z(q)} \end{bmatrix},  \begin{bmatrix} \check{\mbf{c}}^{Z(q)} \\ - \check{\mbf{b}}^{Z(q)} \end{bmatrix} \right)\lzon. \label{eq:faultdiagLZ_celim4}
\end{align}

\begin{remark} \rm \label{rem:AFDvariablechange}
By definition of $\check{Z}^{(q)}$, the set in \eqref{eq:faultdiagLZ_celim4} has less generators/lines than the one in \eqref{eq:desc_separationconditionlzline}. By construction, if \eqref{eq:faultdiagLZ_celim4} holds for all $q \in \mathcal{Q}$, then $\seq{\mbf{u}} = \mbf{c}_u + \mbf{G}_u \seq{\bm{\xi}}_u \in \mathcal{S}(\seq{Y}\zspace^{[\cdot]}(\seq{\mbf{u}}),\modelset,\seq{U})$ as desired. Note that this procedure requires a change in the decision variables, in which the new designed sequence is $\seq{\bm{\xi}}_u$. 
\end{remark}

Lemma \ref{thm:desc_separatinginputlzequiv} further reduces the complexity of \eqref{eq:faultdiagLZ_celim4}, without introducing any conservatism, by replacing the constraint of being outside an LZ with that of being outside a zonotope.
\begin{lemma} \label{thm:desc_separatinginputlzequiv} \rm 
Consider $\check{\mbf{M}}^{Z(q)}$, $\check{\mbf{G}}^{Z(q)}$, $\check{\mbf{c}}^{Z(q)}$, $\check{\mbf{S}}^{Z(q)}$, 
$\check{\mbf{A}}^{Z(q)}$, $\check{\mbf{b}}^{Z(q)}$, $\check{\mbf{G}}_u^{(q)}$, $\check{\mbf{A}}_u^{(q)}$ as in \eqref{eq:constrainteliminationlzreduced}, with $\check{\mbf{S}}^{Z(q)}=\mbf{0}$ and $\check{\mbf{M}}^{Z(q)}$ full column rank$^{\ref{QRdec}}$. Let 
\begin{align*}
& \mbf{N}^{\dagger(q)} \triangleq - \begin{bmatrix} \check{\mbf{G}}_u^{(q)} \\ \check{\mbf{A}}_u^{(q)} \end{bmatrix}, ~ \mbf{M}^{\dagger(q)} \triangleq \begin{bmatrix} \check{\mbf{M}}^{Z(q)} \\ \mbf{0} \end{bmatrix}, \\ & \mbf{G}^{\dagger(q)} \triangleq \begin{bmatrix} \check{\mbf{G}}^{Z(q)} \\ \check{\mbf{A}}^{Z(q)} \end{bmatrix}, ~ \mbf{c}^{\dagger(q)} \triangleq \begin{bmatrix} \check{\mbf{c}}^{Z(q)} \\ - \check{\mbf{b}}^{Z(q)} \end{bmatrix}.
\end{align*}
$\mbf{M}^{\dagger(q)}$ is also full column rank. Then, there exist matrices $\mbf{M}^{-(q)}$, $\mbf{M}^{+(q)}$, $\mbf{G}^{-(q)}$, $\mbf{G}^{+(q)}$, $\mbf{N}^{-(q)}$, $\mbf{N}^{+(q)}$, and vectors $\mbf{c}^{-(q)}$, $\mbf{c}^{+(q)}$, so that
\begin{align}
\eqref{eq:faultdiagLZ_celim4} \iff \mathring{\mbf{N}}^{(q)} \zseq{\bm{\xi}}_u \notin \mathring{\mathcal{Y}}^{(q)} \triangleq ( \mathring{\mbf{G}}^{(q)}, \mathring{\mbf{c}}^{(q)} )\zon, \label{eq:separationconditionlz}
\end{align}
with $\mathring{\mbf{G}}^{(q)} \triangleq \mbf{G}^{-(q)} - \mbf{M}^{-(q)}\inv{(\mbf{M}^{+(q)})} \mbf{G}^{+(q)}$, $\mathring{\mbf{c}}^{(q)} \triangleq \mbf{c}^{-(q)} - \mbf{M}^{-(q)}\inv{(\mbf{M}^{+(q)})} \mbf{c}^{+(q)}$, $\mathring{\mbf{N}}^{(q)} \triangleq \mbf{N}^{-(q)} - \mbf{M}^{-(q)}\inv{(\mbf{M}^{+(q)})} \mbf{N}^{+(q)}$.
\end{lemma}

\proof 
Recall that $\eqref{eq:faultdiagLZ_celim4} \iff \nexists (\bm{\delta}^Z,\bm{\xi}^Z) \in \realset^{n_\delta^Z} \times B_\infty^{n_g^Z} : \mbf{M}^{\dagger(q)} \bm{\delta}^Z + \mbf{G}^{\dagger(q)} \bm{\xi}^Z + \mbf{c}^{\dagger(q)} = \mbf{N}^{\dagger(q)} \zseq{\bm{\xi}}_u$. Since $\mbf{M}^{\dagger(q)}$ is full column rank, then by rearranging the rows of $\mbf{M}^{\dagger(q)} \bm{\delta} + \mbf{G}^{\dagger(q)} \bm{\xi} + \mbf{c}^{\dagger(q)} = \mbf{N}^{\dagger(q)} \zseq{\bm{\xi}}_u$, there exists an invertible matrix $\mbf{M}^{+(q)}$ satisfying
\begin{equation} \label{eq:lemmazonotopeequiv1}
\begin{bmatrix} \mbf{M}^{+(q)} \\ \mbf{M}^{-(q)} \end{bmatrix} \bm{\delta}^Z + \begin{bmatrix} \mbf{G}^{+(q)} \\ \mbf{G}^{-(q)} \end{bmatrix} \bm{\xi}^Z + \begin{bmatrix} \mbf{c}^{+(q)} \\ \mbf{c}^{-(q)} \end{bmatrix} = \begin{bmatrix} \mbf{N}^{+(q)} \\ \mbf{N}^{-(q)} \end{bmatrix} \zseq{\bm{\xi}}_u.
\end{equation}
Therefore, $\mbf{M}^{+(q)} \bm{\delta}^Z + \mbf{G}^{+(q)} \bm{\xi}^Z + \mbf{c}^{+(q)} = \mbf{N}^{+(q)} \zseq{\bm{\xi}}_u$ holds, implying that $\bm{\delta}^Z = \inv{(\mbf{M}^{+(q)})}(- \mbf{G}^{+(q)} \bm{\xi}^Z - \mbf{c}^{+(q)} + \mbf{N}^{+(q)} \zseq{\bm{\xi}}_u)$. Substituting the latter in the lower part of \eqref{eq:lemmazonotopeequiv1} leads to
\begin{align}
& (\mbf{G}^{-(q)} - \mbf{M}^{-(q)}\inv{(\mbf{M}^{+(q)})} \mbf{G}^{+(q)}) \bm{\xi}^Z + (\mbf{I} - \mbf{M}^{-(q)}\inv{(\mbf{M}^{+(q)})}) \mbf{c}^{+(q)} \nonumber\\ & = (\mbf{N}^{-(q)} - \mbf{M}^{-(q)}\inv{(\mbf{M}^{+(q)})} \mbf{N}^{+(q)}) \zseq{\bm{\xi}}_u. \label{eq:lemmazonotopeequiv2}
\end{align}
Therefore $\eqref{eq:faultdiagLZ_celim4} \iff \nexists \bm{\xi}^Z \in B_\infty^{n_g^Z} : \eqref{eq:lemmazonotopeequiv2}$, proving the lemma.\qed

By Lemma \ref{thm:desc_separatinginputlzequiv}, we have $\eqref{eq:separationconditionlz} \implies \seq{\mbf{u}} \in \mathcal{S}(\zseq{Y}\lzon^{[\cdot]}(\seq{\mbf{u}}),\modelset,\seq{U})$. Similarly to the LP in Lemma \ref{thm:separationlzLP}, for each $q \in \mathcal{Q}$, the condition \eqref{eq:separationconditionlz} can be verified for one $\zseq{\bm{\xi}}_u \in B(\zseq{\mbf{A}}_u,\zseq{\mbf{b}}_u)$ by solving the LP
\begin{equation} \label{eq:separationconditionlzLPxiu}
\hat{\kappa}^{(q)}(\seq{\bm{\xi}}_u) \triangleq \left\{ \underset{\kappa^{(q)}, \bm{\xi}^{(q)}}{\min} \kappa^{(q)} : \left. \begin{aligned} & \mathring{\mbf{N}}^{(q)} \seq{\bm{\xi}}_u = \mathring{\mbf{G}}^{(q)} \bm{\xi}^{(q)} + \mathring{\mbf{c}}^{(q)}, \\ & \ninf{\bm{\xi}^{(q)}} \leq 1 + \kappa^{(q)} \end{aligned} \right. \right\},
\end{equation}
where \eqref{eq:separationconditionlz} holds iff $\hat{\kappa}^{(q)}(\seq{\bm{\xi}}_u) > 0$.

\subsection{Input optimization: mixed-integer program formulation}

Consider the design of an optimal input $\seq{\mbf{u}}$ minimizing a functional $J(\seq{\mbf{u}})$ subject to $\seq{\mbf{u}} \in \mathcal{S}(\seq{Y}\zspace^{[\cdot]}(\seq{\mbf{u}}),\modelset,\seq{U})$ for known initial conditions $X_0$ and $\mbf{u}_0$. By rewriting the problem in terms of the variables $\seq{\bm{\xi}}_u$, and using \eqref{eq:separationconditionlz}-\eqref{eq:separationconditionlzLPxiu}, we have that $\seq{\mbf{u}} = \zseq{\mbf{c}}_u + \zseq{\mbf{G}}_u \zseq{\bm{\xi}}_u^* \in \mathcal{S}(\seq{Y}\zspace^{[\cdot]}(\seq{\mbf{u}}),\modelset,\seq{U})$, where $\zseq{\bm{\xi}}_u^*$ is the solution to the optimization problem (the derivation is similar to \cite{Scott2014})
\begin{align}
& \underset{\seq{\bm{\xi}}_u, \kappa^{(q)}, \bm{\xi}^{Z(q)}, \mbf{\lambda}^{(q)}, \bm{\mu}_1^{(q)}, \bm{\mu}_2^{(q)}, \mbf{p}_1^{(q)}, \mbf{p}_2^{(q)}}{\min} ~ J(\seq{\bm{\xi}}_{u}) \label{eq:fault_optimalseparatinglzfinal} \\
\text{s.t.}	~~ &  \seq{\bm{\xi}}_u \in B_\infty(\seq{\mbf{A}}_u, \seq{\mbf{b}}_u), \quad \mbf{K} \zseq{\mbf{G}}_u \zseq{\bm{\xi}}_u = \mbf{u}_0 - \mbf{K} \zseq{\mbf{c}}_u,  \nonumber\\
& \varepsilon \leq \hat{\kappa}^{(q)} \leq \hat{\kappa}_\text{m}^{(q)}, \quad
\mathring{\mbf{N}}^{(q)} \seq{\bm{\xi}}_u = \mathring{\mbf{G}}^{(q)} \bm{\xi}^{Z(q)} + \mathring{\mbf{c}}^{(q)}, \nonumber\\ & \ninf{\bm{\xi}^{Z(q)}} \leq 1 + \kappa^{(q)}, \quad
(\mathring{\mbf{G}}^{(q)})^T \bm{\lambda}^{(q)} = \bm{\mu}_1^{(q)} - \bm{\mu}_2^{(q)}, \nonumber\\
& 1 = (\bm{\mu}_1^{(q)} + \bm{\mu}_2^{(q)})^T \bm{1}, \quad \bm{0} \leq \bm{\mu}_1^{(q)}, \bm{\mu}_2^{(q)}, \nonumber\\
& \mbf{p}_1^{(q)}, \mbf{p}_2^{(q)} \in \{0,1\}^{n_g^{(q)}}, \quad \bm{\mu}_1^{(q)} \leq \mbf{p}_1^{(q)}, \quad \bm{\mu}_2^{(q)} \leq \mbf{p}_2^{(q)}, \nonumber\\
& \xi_j^{Z(q)} - 1 - \kappa^{(q)} \in [-2(1 + \hat{\kappa}_\text{m}^{(q)})(1 - p_{1,j}^{(q)}),\, 0], \nonumber\\
& \xi_j^{Z(q)} + 1 + \kappa^{(q)} \in [0,\, 2(1 + \hat{\kappa}_\text{m}^{(q)})(1 - p_{2,j}^{(q)})], \nonumber
\end{align}
$\forall q  = 1,\ldots,n_q$, with $\varepsilon > 0$ being the \emph{minimum separation threshold}, $\mbf{K} \triangleq [\eye{n_u}\,\; \zeros{n_u}{Nn_u}]$, and $\hat{\kappa}_\text{m}^{(q)}$ is an upper bound for $\hat{\kappa}^{(q)}(\seq{\bm{\xi}}_u)$ defined as in \eqref{eq:separationconditionlzLPxiu}, for all $\seq{\bm{\xi}}_u \in (\seq{\mbf{A}}_u,\seq{\mbf{b}}_u)$.

To minimize the adverse effect of injecting $\seq{\mbf{u}}$ into \eqref{eq:desc_system} relative to a reference input sequence\footnote{This sequence may be generated by a nominal controller, or be a replication of the initial condition $\mbf{u}_0$, for instance.} $\tilde{\mbf{u}} \in \seq{U}$, one may consider the cost $J(\seq{\mbf{u}}) \triangleq (\seq{\mbf{u}} - \tilde{\mbf{u}})\zspace^T \mbf{R} (\seq{\mbf{u}} - \tilde{\mbf{u}})$, where $\mbf{R} \in \realsetmat{(N+1)n_u}{(N+1)n_u}$ is positive definite. Then, in the new variables, $J(\seq{\bm{\xi}}_{u}) = (\seq{\mbf{c}}_u +  \seq{\mbf{G}}_u \seq{\bm{\xi}}_u - \tilde{\mbf{u}})\zspace^T \mbf{R} (\seq{\mbf{c}}_u +  \seq{\mbf{G}}_u \seq{\bm{\xi}}_u - \tilde{\mbf{u}}) = \zseq{\bm{\xi}}_u^T \zseq{\mbf{G}}_u^T \mbf{R} \seq{\mbf{G}}_u \seq{\bm{\xi}}_u + 2(\zseq{\mbf{c}}_u^T \mbf{R} \zseq{\mbf{G}}_u - \tilde{\mbf{u}}^T \mbf{R} \zseq{\mbf{G}}_u) \seq{\bm{\xi}}_u - 2 \zseq{\mbf{c}}_u^T \mbf{R} \tilde{\mbf{u}} + \zseq{\mbf{c}}_u^T \mbf{R} \seq{\mbf{c}}_u + \tilde{\mbf{u}}^T \mbf{R} \tilde{\mbf{u}}$. In this work, we employ a similar approach but with lower computational cost: we use $J(\seq{\mbf{u}}) = \|\mbf{R} (\seq{\mbf{u}}-\tilde{\mbf{u}})\|_1$, which, by introducing $(N+1)n_u$ auxiliary decision variables, results in a mixed-integer linear program (MILP).

\section{Numerical examples} \label{sec:results}

\subsection{State estimation} \label{sec:resultsestimation}

This section compares the results of state estimation using the methods proposed in \cite{Puig2018}\footnote{We use the approach  in \cite{Puig2018} with Kalman correction matrix.}, \cite{Rego2020b}, and Section \ref{sec:desc_estimationLZ} for descriptor systems. These methods use zonotopes, constrained zonotopes, and line zonotopes, respectively, and are referred to as Zonotope, CZ, and LZ. Consider system \eqref{eq:desc_system} with matrices $\mbf{E} = \text{diag}(1,1,0)$, $\mbf{B}_w = \text{diag}(0.1,1.5,0.6)$, $\mbf{D}_v = \text{diag}(0.5,1.5)$,
\begin{equation*}
\mbf{A} = \begin{bmatrix} 0.5 & 0 & 0.5 \\ 0.8 & 0.95 & 0 \\ -1 & 0.5 & 1 \end{bmatrix}, \; \mbf{B} = \begin{bmatrix} 1 & 0 \\ 0 & 1 \\ 0 & 0 \end{bmatrix}, \; \mbf{C} = \begin{bmatrix} 1 & 0 & 1 \\ 1 & -1 & 0 \end{bmatrix}, 
\end{equation*}
and $\mbf{D} = \mbf{0}$. The known input is $\mbf{u}_k = (0.5 \sin(kT_s) + 1, -2 \cos(kT_s))$, with $T_s$ $= 0.1 \pi$s. The initial  $\mbf{x}_0$ is bounded by 
\begin{equation} \label{eq:desc_estimationx0cz}
X_0 = (\text{diag}(0.1, 1.5, 0.6),\, [ 0.5 \; 0.5 \; 0.25]^T)\zon,
\end{equation}
with uncertainties modeled as uniformly distributed random disturbances bounded by $\|\mbf{w}_k\|_\infty \leq 1$, $\|\mbf{v}_k\|_\infty \leq 1$. To demonstrate the capabilities of the LZ method, the considered LZ initial set is the entire state space $X_0=\realset^3$. The CZ $X_\text{A} \subset \realset^3$ in Assumption 1 in \cite{Rego2020b}, i.e., satisfying $\mbf{x}_k \in X_\text{A}, \forall k\geq 0$, is chosen as $X_\text{A} = \{50{\cdot}\eye{3},\mbf{0}\}$. The simulation is conducted for $\mbf{x}_0 = [ 0.5 \; 0.5 \; 0.25 ]^T$, using Gurobi 10.0.1 and MATLAB 9.1, on a laptop with 32 GB RAM and an Intel Core i7-12700H processor. The complexity of CZs and LZs  is limited to 30 generators and 5 constraints using the constraint elimination algorithm and generator reduction both described in \cite{Scott2016}. The complexity of the zonotopes is limited to 30 generators using Method 4 in \cite{Scott2018}. For LZs, lines are eliminated prior to eliminating constraints, as proposed in Section \ref{sec:desc_lzcomplexityreduction}. 
\begin{figure}[!tb]
	\centering{
		\def\svgwidth{1\columnwidth}
		{\scriptsize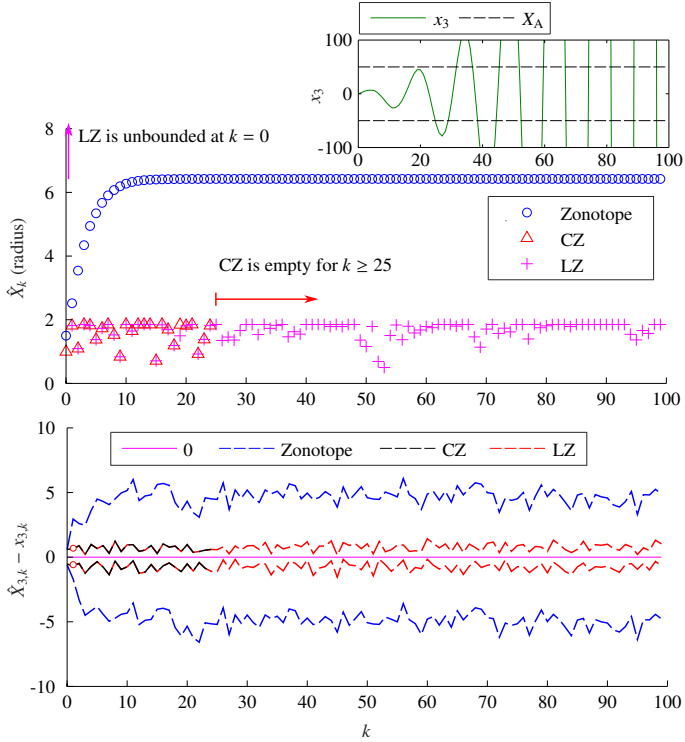}
		\caption{The radii of the enclosures $\hat{X}_k$ obtained using the methods proposed in \cite{Puig2018}, \cite{Rego2020b}, and the line zonotope method proposed in Section \ref{sec:desc_estimationLZ} (top), as well as the projections of $\hat{X}_k$ onto $x_3$ centered at $x_3$ (bottom). The circles at the bottom red curves denote the instant $k$ in which the enclosure generated by LZ becomes bounded.}\label{fig:desc_lzestimation}}
\end{figure}

Figure \ref{fig:desc_lzestimation} shows the radii of the enclosures $\hat{X}_k$ for $k \in [0,100]$ obtained using Zonotope, CZ, and LZ, as well as the projections of $\hat{X}_k$ onto $x_3$, centered at $x_3$. As it can be noticed, both CZ and LZ provide substantially sharper bounds in comparison to zonotopes. This is possible since the enclosures proposed in \cite{Rego2020b} and Lemma \ref{lem:desc_predictionlz} explicitly take the static constraints into account. However, since the system is unstable, the state $\mbf{x}_k$ does not lie in $X_\text{A}$ for the entire simulation, leading to empty sets in CZ for $k \geq 25$. In contrast, LZ is able to provide enclosures that are as accurate as the ones obtained using CZ, but without leading to empty sets since it does not require the existence of an admissible set $X_\text{A}$. Moreover, the enclosure $\hat{X}_k$ provided by LZ at $k=0$ is unbounded, as it assumes the entire state space as initial set for demonstration purposes. Thanks to the measurements, which refine the computed sets, this enclosure becomes bounded at $k=1$ (as shown by the circles on the red curves in the lower figure). It is important to note that although $X_\text{A}$ can be chosen arbitrarily large, the unstable trajectory of the system will eventually exit $X_\text{A}$ at a given $k \geq 0$, rendering the CZ method ineffective in this scenario. Conversely, the LZ method was able to construct the sets in just a few steps without requiring a bounded $X_\text{A}$. The average computational times for Zonotope, CZ, and LZ, over $k \in [1, 21]$, are 0.37 ms, 2.43 ms, and 2.02 ms, respectively. While the Zonotope method is significantly faster, it results in considerable conservatism. In contrast, LZ outperforms CZ in terms of speed because line elimination has a lower computational complexity than generator reduction (see Section \ref{sec:desc_lzcomplexityreduction}). Notably, when the LZs become bounded, both methods eliminate the same variables, with the key difference being that LZ eliminates lines instead of generators from $X_\text{A}$, further demonstrating the advantages of the new set representation. 

\subsection{Active fault diagnosis} \label{sec:desc_examplesAFDCZ}

We now evaluate the effectiveness of the AFD method proposed in Section \ref{sec:AFDLZ}, denoted by AFDLZ. We compare the results with the CZ method described in \ref{app:AFDCZ}, denoted by AFDCZ\footnote{As it shown in Section \ref{sec:resultsestimation}, and also in \cite{Rego2020b}, zonotopes provide very weak enclosures for LDS in comparison to CZs and LZs. For this reason, we do not conduct comparisons with zonotope methods in this section.}. Both AFDLZ and AFDCZ rely on reachable tubes and therefore are expected to outperform \cite{Rego2020b} which relies on the final reachable set only, albeit at the cost of increased computational complexity\footnote{Since the focus is on highlighting the advantages of unbounded sets, a direct comparison with \cite{Rego2020b} is omitted. However, a similar comparison can be found in \cite{Scott2014} for the linear case.}. 
The input sequences designed by these methods are denoted by $\seq{\mbf{u}}_\text{AFDLZ}$ and $\seq{\mbf{u}}_\text{AFDCZ}$, respectively. Consider \eqref{eq:desc_system} with $\mbf{E}^{[1]}$, $\mbf{A}^{[2]}$, $\mbf{B}^{[1]}$, and $\mbf{B}_w^{[1]}$ as in the previous example (Section \ref{sec:resultsestimation}), and $\mbf{C}^{[1]} = [ 1 \,\; -1 \,\; 0 ]$, $\mbf{C}^{[4]} = [ 1 \,\; -1 \,\; 0.1]$,
\begin{equation*}
\mbf{A}^{[1]} = \begin{bmatrix} 0.5 & 0 & 0 \\ 0.8 & 0.95 & 0 \\ -1 & 0.5 & 1 \end{bmatrix}, \; \mbf{B}^{[3]} = \begin{bmatrix} 1 & 0 \\ 0 & 0.9 \\ -0.1 & 0 \end{bmatrix}, 
\end{equation*}
$\mbf{D}_v^{[1]} = 0.5$, $\mbf{E}^{[i]} = \mbf{E}^{[1]}$, $\mbf{A}^{[3]} = \mbf{A}^{[4]} = \mbf{A}^{[1]}$, $\mbf{B}^{[2]} = \mbf{B}^{[4]} = \mbf{B}^{[1]}$, $\mbf{C}^{[2]} = \mbf{C}^{[3]} = \mbf{C}^{[1]}$, and $\mbf{D}_v^{[i]} = \mbf{D}_v^{[1]}$, $\mbf{D}^{[i]} = \mbf{D}^{[1]} = \zeros{1}{2}$, $i \in \{2,3,4\}$. The initial state $\mbf{x}_0^{[i]}$ is bounded by \eqref{eq:desc_estimationx0cz}, 
the uncertainties are bounded by $\|\mbf{w}_k\|_\infty \leq 0.1$, $\|\mbf{v}_k\|_\infty \leq 0.1$, and the input is limited by $\|\mbf{u}_k\|_\infty \leq 8$. The initial input is $\mbf{u}_0 = (0.5,-1)$, and the reference sequence is given by $\tilde{\mbf{u}} = (\mbf{u}_0,\ldots,\mbf{u}_0) \in \realset^{2(N+1)}$. For the AFDCZ, we consider $X_\text{A} = \{\eye{3},\mbf{0}\}$. The separation threshold is $\varepsilon = 1 \ten{-4}$. The separation horizon is chosen as $N=3$. All the models $i \in \modelset$ are considered to be faulty and must be separated. The number of generators in AFDLZ and AFDCZ is limited to 1.6 times the dimension of the sets, with the number of constraints limited to 2, while the number of lines in AFDLZ is reduced to the minimum possible. The optimal input sequences were obtained using Gurobi 10.0.1 and MATLAB 9.1, with $J(\seq{\mbf{u}}) = \|\seq{\mbf{u}} - \tilde{\mbf{u}} \|_1$, and are
\begin{equation*}
\begin{aligned}
\seq{\mbf{u}}_\text{AFDLZ} & = 
\left(\begin{bmatrix} 0.5 \\ -1 \end{bmatrix},\begin{bmatrix} -3.4399 \\ -1 \end{bmatrix}, \begin{bmatrix} 1.0854 \\ -7.0632 \end{bmatrix}, \begin{bmatrix} 0.5 \\ -1 \end{bmatrix} \right), \\
\seq{\mbf{u}}_\text{AFDCZ} & =
\left(\begin{bmatrix} 0.5 \\ -1 \end{bmatrix},\begin{bmatrix} 2.9286 \\ -1 \end{bmatrix}, \begin{bmatrix} 3.1628 \\ -1 \end{bmatrix}, \begin{bmatrix} 0.5 \\ -1  \end{bmatrix} \right). 
\end{aligned}
\end{equation*}

\begin{figure}[!tb]
	\centering{
		\def\svgwidth{0.9\columnwidth}
		{\scriptsize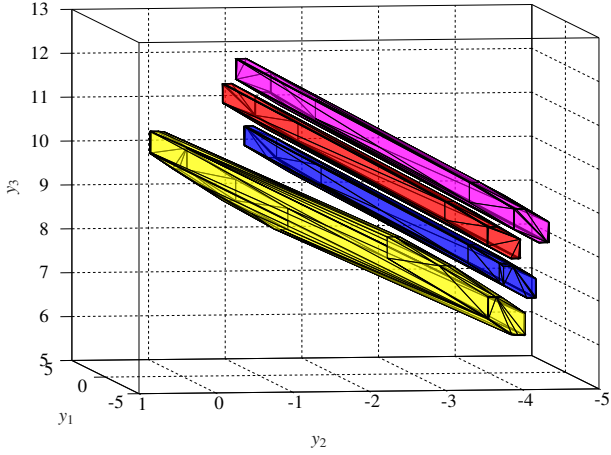}
		\caption{Projections of the ORTs $\zseq{Y}\lzon^{[i]}(\zseq{\mbf{u}}_\text{AFDLZ})$ onto $k \in \{1,2,3\}$, for models $i=1$ (red), $i=2$ (yellow), $i=3$ (blue), and $i=4$ (magenta).}\label{fig:desc_lzafd}}
\end{figure}

For a comparison analysis, we compute $\zseq{Y}\lzon^{[i]}$ for all $i \in \modelset$, obtained by injecting $\tilde{\mbf{u}}$, $\seq{\mbf{u}}_\text{AFDCZ}$, and $\seq{\mbf{u}}_\text{AFDLZ}$. Table \ref{tab:intersection} shows $\zseq{Y}\lzon^{[i]} \cap \zseq{Y}\lzon^{[j]}$ for $i \neq j$, where $*$ denotes a non-empty intersection, and $\cmrk$ being an empty intersection. The reference sequence $\tilde{\mbf{u}}$ does not separate any of the ORTs (omitted from the table), while $\seq{\mbf{u}}_\text{AFDCZ}$ is able to separate only a few combinations. The latter is due to the fact that one of the models in $\modelset$ is unstable, leading to a violation of Assumption 1 in \cite{Rego2020b}. On the other hand, $\seq{\mbf{u}}_\text{AFDLZ}$ separates all the ORTs, leading to guaranteed fault diagnosis. 
Additionally, we generate 500 samples of $\zseq{\mbf{y}}^{[i]}$ for each $i \in \modelset$ using $\seq{\mbf{u}}_\text{AFDCZ}$ and $\seq{\mbf{u}}_\text{AFDLZ}$, and verify the inclusion $\zseq{\mbf{y}}^{[i]} \in \zseq{Y}\lzon^{[j]}$ for all $i,j \in \modelset$ (this inclusion can be verified through an LP similar to Proposition 2 in \cite{Scott2016}). As shown in Table \ref{tab:samples}, in contrast to the other input sequences, $\seq{\mbf{u}}_\text{AFDLZ}$ is able to separate all the cases, with $\zseq{\mbf{y}}^{[i]} \in \zseq{Y}\lzon^{[j]}$ holding only for $i=j$. Finally, %
Figure \ref{fig:desc_lzafd} shows the projections of the ORTs $\zseq{Y}\lzon^{[i]}(\zseq{\mbf{u}}_\text{AFDLZ})$ onto $k \in \{1,2,3\}$ for $i \in \{1,2,3,4\}$, resulting from the injection of the designed $\zseq{\mbf{u}}_\text{AFDLZ}$. Note that $\zseq{Y}\lzon^{[i]}(\zseq{\mbf{u}}_\text{AFDLZ})$ are disjoint for every $i \in \modelset$, showing that the injection of $\zseq{\mbf{u}}_\text{AFDLZ}$ guarantees fault diagnosis in the given time interval. Furthermore, although $X_\text{A}$ can be chosen arbitrarily large, there is no guarantee that it will encompass the system trajectories in all faulty scenarios, particularly since model $i=2$ is unstable. As a result, AFDCZ is unsuitable for this case, as fault diagnosis cannot be ensured. Finally, the total computational time for set complexity reduction was 303.1 ms for CZs and 12.5 ms for LZs. This highlights the computational advantage of using LZs for AFD in linear descriptor systems, demonstrating that line elimination can be significantly more efficient than standard order reduction.

\begin{table}[!tb] 
  \scriptsize
    \centering
	\caption{Intersection of ORTs:  $\cmrk$ if empty, $*$ otherwise.}
	\begin{tabular}{c c c c | c c c c} \hline
		\multicolumn{4}{c|}{$\zseq{\mbf{u}}_\text{AFDLZ}$} & \multicolumn{4}{c}{$\zseq{\mbf{u}}_\text{AFDCZ}$} \\ \hline
		$\cap$ & $\zseq{Y}\lzon^{[2]}$ & $\zseq{Y}\lzon^{[3]}$ & $\zseq{Y}\lzon^{[4]}$ & $\cap$ & $\zseq{Y}\lzon^{[2]}$ & $\zseq{Y}\lzon^{[3]}$ & $\zseq{Y}\lzon^{[4]}$ \\ \hline
	    $\zseq{Y}\lzon^{[1]}$ & $\cmrk$ & $\cmrk$ & $\cmrk$ & $\zseq{Y}\lzon^{[1]}$ & $\cmrk$ & $*$ & $*$ \\
	    $\zseq{Y}\lzon^{[2]}$ &   & $\cmrk$ & $\cmrk$ & $\zseq{Y}\lzon^{[2]}$ &   & $\cmrk$ & $\cmrk$ \\
	    $\zseq{Y}\lzon^{[3]}$ &   &   & $\cmrk$ & $\zseq{Y}\lzon^{[3]}$ &   &   & $*$ \\
		\hline
	\end{tabular} \normalsize
 	\label{tab:intersection}
\end{table}

\begin{table}[!tb] 
  \scriptsize
    \centering
	\caption{Inclusion $\zseq{\mbf{y}}^{[i]} \in \zseq{Y}\lzon^{[j]}$ for all $i,j \in \modelset$, considering 500 samples of $\zseq{\mbf{y}}^{[i]}$}.
	\begin{tabular}{c c c c c | c c c c c} \hline
		\multicolumn{5}{c|}{$\zseq{\mbf{u}}_\text{AFDLZ}$} & \multicolumn{5}{c}{$\zseq{\mbf{u}}_\text{AFDCZ}$} \\ \hline
		$\in$ & $\zseq{Y}\lzon^{[1]}$ & $\zseq{Y}\lzon^{[2]}$ & $\zseq{Y}\lzon^{[3]}$ & $\zseq{Y}\lzon^{[4]}$ & $\in$ &
		$\zseq{Y}\lzon^{[1]}$ & $\zseq{Y}\lzon^{[2]}$ & $\zseq{Y}\lzon^{[3]}$ & $\zseq{Y}\lzon^{[4]}$ \\ \hline
	    $\zseq{\mbf{y}}^{[1]}$ & 500 &   0 &   0 &   0 & 
	    $\zseq{\mbf{y}}^{[1]}$ & 500 &   0 & 483 & 311 \\
	    $\zseq{\mbf{y}}^{[2]}$ &   0 & 500 &   0 &   0 & 
	    $\zseq{\mbf{y}}^{[2]}$ &   0 & 500 &   0 &   0\\
	    $\zseq{\mbf{y}}^{[3]}$ &   0 &   0 & 500 &   0 & 
	    $\zseq{\mbf{y}}^{[3]}$ & 479 &   0 & 500 & 98\\
        $\zseq{\mbf{y}}^{[4]}$ &   0 &   0 &   0 & 500 & 
        $\zseq{\mbf{y}}^{[4]}$ & 377 &   0 & 155 & 500 \\
		\hline
	\end{tabular} \normalsize
 	\label{tab:samples}
\end{table}

We now investigate the performance of AFDLZ for a system with unbounded initial set $X_0$. Consider a system whose dynamics obey one of possible $n_m$ known models $x_k^{[i]} = a^{[i]}  x_{k-1}^{[i]} + b^{[i]} u_{k-1} + b_w^{[i]}  w_{k-1}$, and $y_k^{[i]} = c^{[i]}  x_k^{[i]} + d^{[i]} u_{k} + d_v^{[i]} v_{k}$, where $a^{[1]} = 1$, $a^{[2]} = 0.2$, $a^{[3]} = -0.5$, $b^{[i]} = 1$, $b_w^{[i]} = 1$,  $c^{[i]} = 1$, $d^{[i]} = 0$, $d_v^{[i]} = 1$, $i \in \{1,2,3\}$. The initial set is $x_0^{[i]} \in \realset = (1,0,0)\lzon$. The uncertainties are bounded by $|w_k| \leq 0.1$, $|v| \leq 0.1$, and the input is limited by $|u_k| \leq 2$. The separation threshold is $\varepsilon = 1 \ten{-1}$. The initial input is $u_0 = 0$, and the reference sequence is given by $\tilde{\mbf{u}} = \zeros{(N+1)}{1}$. The separation horizon is chosen as $N=3$. All the models $i \in \modelset$ are considered to be faulty and must be separated. The number of generators in AFDLZ is limited to 3 times the dimension of the sets. The separating input sequence was obtained for $J(\seq{\mbf{u}}) = \|\seq{\mbf{u}}\|_1$, and is $\seq{\mbf{u}}_\text{AFDLZ} = (0,0.99,0,0)$. %
We compute $\zseq{Y}\lzon^{[i]}$ for all $i \in \modelset$, obtained by injecting $\tilde{\mbf{u}}$ and $\seq{\mbf{u}}_\text{AFDLZ}$, and verify $\zseq{Y}\lzon^{[i]} \cap \zseq{Y}\lzon^{[j]}$ for $i \neq j$. As in the previous case, the reference sequence $\tilde{\mbf{u}}$ does not separate any of $\zseq{Y}\lzon^{[i]}$. On the other hand, $\seq{\mbf{u}}_\text{AFDLZ}$ separates all the unbounded ORTs, leading to guaranteed AFD even if no bounds were known for the initial state. A three-dimensional view of the unbounded ORTs can be found in \url{https://youtu.be/LKIcve0zSAw}.

\section{Conclusions}
\label{sec:conclusions}

This paper introduced new methods for set-based state estimation and AFD of LDS with bounded uncertainties, which leverage line zonotopes (LZs) -- a novel set representation that extends convex zonotopes (CZs) to unbounded sets while retaining most of their computational benefits, as well as similar complexity reduction methods. This enabled characterizing reachable sets in unstable and unobservable LDS without requiring a bounded enclosure for all $k\geq 0$. We also proposed a tube-based AFD approach that exploits the entire output sequence, enabling earlier fault isolation even in unbounded cases while reducing input sequence length and energy. Our methods outperform CZ-based approaches in challenging scenarios, as demonstrated through numerical examples. Future work will investigate alternative approaches for optimal input design that do not require solving a mixed-integer program.










\appendix

\section{Fault diagnosis: state reachable tubes} \label{sec:app_statetube}

The SRT $\zseq{Z}\lzon^{[i]}(\seq{\mbf{u}})$ obtained in Section \ref{sec:AFDLZ} is given by $\zseq{Z}\lzon^{[i]}(\seq{\mbf{u}}) = (\seq{\mbf{M}}\zerospace^{[i]}, \zseq{\mbf{G}}^{[i]}, \zseq{\mbf{c}}^{[i]}(\seq{\mbf{u}}), \zseq{\mbf{S}}^{[i]}, \zseq{\mbf{A}}^{[i]}, \zseq{\mbf{b}}^{[i]}(\seq{\mbf{u}}) )\lzon$, where  $\zseq{\mbf{c}}^{[i]}(\seq{\mbf{u}}) = \zseq{\mbf{Q}}^{[i]} \mbf{c}_\sigma + \zseq{\mbf{p}}^{[i]} + \zseq{\mbf{H}}^{[i]} \seq{\mbf{u}}$, $\zseq{\mbf{M}}^{[i]} = [ \zseq{\mbf{Q}}^{[i]} \mbf{M}_\sigma^{[i]} \,\; \zseq{\mbf{P}}^{[i]}_\text{M}]$, $ \zseq{\mbf{G}}^{[i]} = [ \zseq{\mbf{Q}}^{[i]} \mbf{G}_\sigma^{[i]} \,\; \zseq{\mbf{P}}^{[i]}_\text{G}]$, $\zseq{\mbf{b}}^{[i]}(\seq{\mbf{u}}) = \zseq{\bm{\alpha}}^{[i]} + \zseq{\bm{\Lambda}}^{[i]} \mbf{c}_\sigma + \zseq{\bm{\Omega}}^{[i]} \seq{\mbf{u}}$, and
\begin{align}
\zseq{\mbf{A}}^{[i]} & = \begin{bmatrix} \begin{bmatrix} \mbf{A}_\sigma \\ \check{\mbf{A}}_z^{[i]} \mbf{G}_\sigma^{[i]} \end{bmatrix} & \begin{bmatrix} \mbf{0} \\ \mbf{0} \end{bmatrix} & \cdots & \begin{bmatrix} \mbf{0} \\ \mbf{0} \end{bmatrix} \\
\check{\mbf{A}}_z^{[i]} \begin{bmatrix} \tilde{\mbf{A}}_z^{[i]} \\ \mbf{0} \end{bmatrix} \mbf{G}_\sigma^{[i]} & \check{\mbf{A}}_z^{[i]} \begin{bmatrix} \mbf{0} \\ \mbf{G}_\text{A} \end{bmatrix} & \cdots & \begin{bmatrix} \mbf{0} \\ \mbf{0} \end{bmatrix} \\
\check{\mbf{A}}_z^{[i]} \begin{bmatrix} \tilde{\mbf{A}}_z^{[i]} \\ \mbf{0} \end{bmatrix}^2 \mbf{G}_\sigma{[i]} & \check{\mbf{A}}_z^{[i]} \begin{bmatrix} \tilde{\mbf{A}}_z^{[i]} \\ \mbf{0} \end{bmatrix} \begin{bmatrix} \mbf{0} \\ \mbf{G}_\text{A} \end{bmatrix}  & \cdots & \begin{bmatrix} \mbf{0} \\ \mbf{0} \end{bmatrix} \\
\vdots & \vdots & \ddots & \vdots \\
\check{\mbf{A}}_z^{[i]} \begin{bmatrix} \tilde{\mbf{A}}_z^{[i]} \\ \mbf{0} \end{bmatrix}^N \mbf{G}_\sigma^{[i]} & \check{\mbf{A}}_z^{[i]} \begin{bmatrix} \tilde{\mbf{A}}_z^{[i]} \\ \mbf{0} \end{bmatrix}^{N-1} \begin{bmatrix} \mbf{0} \\ \mbf{G}_\text{A} \end{bmatrix}  & \cdots & \check{\mbf{A}}_z^{[i]} \begin{bmatrix} \mbf{0} \\ \mbf{G}_\text{A} \end{bmatrix} \\
\mbf{0} & \mbf{A}_\text{A} & \cdots & \mbf{0} \\
\mbf{0} & \mbf{0} & \cdots & \mbf{0} \\
\vdots & \vdots & \ddots & \vdots \\
\mbf{0} & \mbf{0} & \cdots & \mbf{A}_\text{A}
\end{bmatrix},  \label{eq:tubelz_A} \\
\zseq{\mbf{S}}^{[i]} & = \begin{bmatrix} \begin{bmatrix} \mbf{S}_\sigma \\ \check{\mbf{A}}_z^{[i]} \mbf{M}_\sigma^{[i]} \end{bmatrix} & \begin{bmatrix} \mbf{0} \\ \mbf{0} \end{bmatrix} & \cdots & \begin{bmatrix} \mbf{0} \\ \mbf{0} \end{bmatrix} \\
\check{\mbf{A}}_z^{[i]} \begin{bmatrix} \tilde{\mbf{A}}_z^{[i]} \\ \mbf{0} \end{bmatrix} \mbf{M}_\sigma^{[i]} & \check{\mbf{A}}_z^{[i]} \begin{bmatrix} \mbf{0} \\ \mbf{M}_\text{A} \end{bmatrix} & \cdots & \begin{bmatrix} \mbf{0} \\ \mbf{0} \end{bmatrix} \\
\check{\mbf{A}}_z^{[i]} \begin{bmatrix} \tilde{\mbf{A}}_z^{[i]} \\ \mbf{0} \end{bmatrix}^2 \mbf{M}_\sigma^{[i]} & \check{\mbf{A}}_z^{[i]} \begin{bmatrix} \tilde{\mbf{A}}_z^{[i]} \\ \mbf{0} \end{bmatrix} \begin{bmatrix} \mbf{0} \\ \mbf{M}_\text{A} \end{bmatrix}  & \cdots & \begin{bmatrix} \mbf{0} \\ \mbf{0} \end{bmatrix} \\
\vdots & \vdots & \ddots & \vdots \\
\check{\mbf{A}}_z^{[i]} \begin{bmatrix} \tilde{\mbf{A}}_z^{[i]} \\ \mbf{0} \end{bmatrix}^N \mbf{M}_z^{[i]} & \check{\mbf{A}}_z^{[i]} \begin{bmatrix} \tilde{\mbf{A}}_z^{[i]} \\ \mbf{0} \end{bmatrix}^{N-1} \begin{bmatrix} \mbf{0} \\ \mbf{M}_\text{A} \end{bmatrix}  & \cdots & \check{\mbf{A}}_z^{[i]} \begin{bmatrix} \mbf{0} \\ \mbf{M}_\text{A} \end{bmatrix} \\
\mbf{0} & \mbf{S}_\text{A} & \cdots & \mbf{0} \\
\mbf{0} & \mbf{0} & \cdots & \mbf{0} \\
\vdots & \vdots & \ddots & \vdots \\
\mbf{0} & \mbf{0} & \cdots & \mbf{S}_\text{A},
\end{bmatrix} \label{eq:tubelz_S}
\end{align} 
with $\zseq{\bm{\alpha}}^{[i]} \triangleq \zseq{\bm{\beta}}^{[i]} + \zseq{\bm{\Upsilon}}^{[i]} \zseq{\mbf{p}}^{[i]}$, $\zseq{\bm{\Lambda}}^{[i]} \triangleq \zseq{\bm{\Upsilon}}^{[i]} \zseq{\mbf{Q}}^{[i]}$, $\zseq{\bm{\Omega}}^{[i]} \triangleq \zseq{\bm{\Gamma}}^{[i]} + \zseq{\bm{\Upsilon}}^{[i]} \zseq{\mbf{H}}^{[i]}$, and the variables
\begin{align}
& \seq{\mbf{p}}\zspace^{[i]} \triangleq \left[ \begin{matrix} \begin{bmatrix} \mbf{0} \\ \mbf{0} \end{bmatrix}^T & \begin{bmatrix} \mbf{0} \\ \mbf{c}_\text{A}^{[i]} \end{bmatrix}^T & \cdots & \sum_{m=1}^{N} \left( \begin{bmatrix} \tilde{\mbf{A}}_z^{[i]} \\ \mbf{0} \end{bmatrix}^{m-1} \begin{bmatrix} \mbf{0} \\ \mbf{c}_\text{A}^{[i]} \end{bmatrix} \right)^T \end{matrix} \right]^T, \label{eq:tubelz_p} \\
& \seq{\mbf{Q}}\zspace^{[i]} \triangleq \left[\begin{matrix} \cdots & \!\!\!\!\left(\begin{bmatrix} \tilde{\mbf{A}}_z^{[i]} \\ \mbf{0} \end{bmatrix}^\ell \right)^T \!\!\!\!\!\!\! & \cdots \end{matrix}\right]^T\!\!, \; \seq{\mbf{H}}\zerospace^{[i]} = \big[\begin{matrix} \cdots & (\mbf{H}_\ell^{[i]})^T & \cdots \end{matrix}\big]^T\!\!, \label{eq:tubelz_QH} \\
& \mbf{H}_\ell^{[i]} \triangleq \left[\begin{matrix} \cdots & \underbrace{\begin{bmatrix} \tilde{\mbf{A}}_z^{[i]} \\ \mbf{0} \end{bmatrix}^{\ell-m} \begin{bmatrix} \tilde{\mbf{B}}^{[i]} \\ \mbf{0} \end{bmatrix}}_{m = 1,2,\ldots,\ell} & \cdots & \underbrace{\begin{bmatrix} \mbf{0} \\ \mbf{0} \end{bmatrix}}_{N-\ell+1 \text{ blocks}} & \cdots \end{matrix} \right], \label{eq:tubelz_Hl}  \\
& \seq{\bm{\beta}}\zspace^{[i]} \triangleq \begin{bmatrix} (\mbf{b}_\sigma^{[i]})^T & \zeros{1}{(N+1)\check{n}_z} & (\mbf{b}_\text{A}^{[i]})^T  & \cdots & (\mbf{b}_\text{A}^{[i]})^T ] \end{bmatrix}^T, \label{eq:tubelz_beta} \\
& \seq{\bm{\Upsilon}}\zspace^{[i]} \triangleq \begin{bmatrix} \zeros{(N+1)(n+n_w)}{n_{c_z}} & \bm{\Xi}( -\check{\mbf{A}}_z^{[i]})^T &  \zeros{(N+1)(n+n_w)}{n_{c_\text{A}}} \end{bmatrix}^T, \label{eq:tubelz_upsilon}\\
& \seq{\bm{\Gamma}}\zspace^{[i]} \triangleq \begin{bmatrix} \zeros{(N+1)n_u}{n_{c_z}} & \bm{\Xi}( -\check{\mbf{B}}^{[i]})^T &  \zeros{(N+1)n_u}{(N+1)n_{c_\text{A}}} \end{bmatrix}^T, \label{eq:tubelz_gamma} \\
& \zseq{\mbf{P}}^{[i]}_\text{G} \triangleq \begin{bmatrix} \begin{bmatrix} \mbf{0} \\ \mbf{0} \end{bmatrix} & \begin{bmatrix} \mbf{0} \\ \mbf{0} \end{bmatrix} & \cdots & \begin{bmatrix} \mbf{0} \\ \mbf{0} \end{bmatrix} \\ \begin{bmatrix} \mbf{0} \\ \mbf{G}_\text{A} \end{bmatrix} & \begin{bmatrix} \mbf{0} \\ \mbf{0} \end{bmatrix} & \cdots & \begin{bmatrix} \mbf{0} \\ \mbf{0} \end{bmatrix} \\
\begin{bmatrix} \tilde{\mbf{A}}_z^{[i]} \\ \mbf{0} \end{bmatrix} \begin{bmatrix} \mbf{0} \\ \mbf{G}_\text{A} \end{bmatrix} & \begin{bmatrix} \mbf{0} \\ \mbf{G}_\text{A} \end{bmatrix} & \cdots & \begin{bmatrix} \mbf{0} \\ \mbf{0} \end{bmatrix} \\
\vdots & \vdots & \ddots & \vdots \\
\begin{bmatrix} \tilde{\mbf{A}}_z^{[i]} \\ \mbf{0} \end{bmatrix}^{N-1}  \begin{bmatrix} \mbf{0} \\ \mbf{G}_\text{A} \end{bmatrix} & \begin{bmatrix} \tilde{\mbf{A}}_z^{[i]} \\ \mbf{0} \end{bmatrix}^{N-2} \begin{bmatrix} \mbf{0} \\ \mbf{G}_\text{A} \end{bmatrix}  & \cdots & \begin{bmatrix} \mbf{0} \\ \mbf{G}_\text{A} \end{bmatrix} \end{bmatrix}, \label{eq:tubelz_PG} \\
& \zseq{\mbf{P}}^{[i]}_\text{M} \triangleq \begin{bmatrix} \begin{bmatrix} \mbf{0} \\ \mbf{0} \end{bmatrix} & \begin{bmatrix} \mbf{0} \\ \mbf{0} \end{bmatrix} & \cdots & \begin{bmatrix} \mbf{0} \\ \mbf{0} \end{bmatrix} \\ \begin{bmatrix} \mbf{0} \\ \mbf{M}_\text{A} \end{bmatrix} & \begin{bmatrix} \mbf{0} \\ \mbf{0} \end{bmatrix} & \cdots & \begin{bmatrix} \mbf{0} \\ \mbf{0} \end{bmatrix} \\
\begin{bmatrix} \tilde{\mbf{A}}_z^{[i]} \\ \mbf{0} \end{bmatrix} \begin{bmatrix} \mbf{0} \\ \mbf{M}_\text{A} \end{bmatrix} & \begin{bmatrix} \mbf{0} \\ \mbf{M}_\text{A} \end{bmatrix} & \cdots & \begin{bmatrix} \mbf{0} \\ \mbf{0} \end{bmatrix} \\
\vdots & \vdots & \ddots & \vdots \\
\begin{bmatrix} \tilde{\mbf{A}}_z^{[i]} \\ \mbf{0} \end{bmatrix}^{N-1}  \begin{bmatrix} \mbf{0} \\ \mbf{M}_\text{A} \end{bmatrix} & \begin{bmatrix} \tilde{\mbf{A}}_z^{[i]} \\ \mbf{0} \end{bmatrix}^{N-2} \begin{bmatrix} \mbf{0} \\ \mbf{M}_\text{A} \end{bmatrix}  & \cdots & \begin{bmatrix} \mbf{0} \\ \mbf{M}_\text{A} \end{bmatrix} \end{bmatrix}, \label{eq:tubelz_PM}
\end{align} 
with $\ell = 0,\ldots,N$. The expression $\mbf{H}_\ell^{[i]}$ holds for $\ell = 1,\ldots,N$, while $\mbf{H}_0^{[i]} = \zeros{n}{(N{+}1)n_u}$. The operator $\bm{\Xi}(\cdot)$ is defined as $\bm{\Xi}(\cdot) \triangleq \text{bdiag}((\cdot),\ldots,(\cdot))$, with $N+1$ matrices $(\cdot)$.

\section{Tube-based AFD of LDS using CZs} \label{app:AFDCZ}

The work in \cite{Rego2020b} has presented a method based on the separation of the final output reachable set to address the problem of AFD of descriptor systems using constrained zonotopes. For comparison with the LZ method developed in Section \ref{sec:AFDLZ}, we develop a tube-based AFD method using CZs allowing for the separation of output tubes in CG-rep along the time interval $k \in [0,N]$. 

Similarly to Section \ref{sec:AFDLZ}, for each model $i \in \modelset$, consider the CZ $Z_\text{A}^{[i]} = \inv{(\mbf{T}^{[i]})} X_\text{A} \times W = ( \mbf{G}_\text{A}^{[i]}, \mbf{c}_\text{A}^{[i]}, \mbf{A}_\text{A}^{[i]}, \mbf{b}_\text{A}^{[i]})_\text{CZ}$, where $X_\text{A}$ is a CZ satisfying Assumption 1 in \cite{Rego2020b}. Moreover, let $ (\mbf{G}_z^{[i]},\mbf{c}_z^{[i]},\mbf{A}_z^{[i]},\mbf{b}_z^{[i]})\czon \triangleq \inv{(\mbf{T}^{[i]})} X_0 \times W $, and define the initial feasible set $Z_0^{[i]} (\mbf{u}_0) = \{\mbf{z} \in \inv{(\mbf{T}^{[i]})} X_0 \times W : \eqref{eq:desc_systemSVDfaultconstraintslz} \text{ holds for }k=0\}$. This set is given by $Z_0^{[i]}(\mbf{u}_0) = (\mbf{G}_0^{[i]},\mbf{c}_0^{[i]},\mbf{A}_0^{[i]},$ $\mbf{b}_0^{[i]}(\mbf{u}_0))\czon$, where $\mbf{G}_0^{[i]} = \mbf{G}_z^{[i]}$, $\mbf{c}_0^{[i]} = \mbf{c}_z^{[i]}$,
\begin{equation} \label{eq:initialAb}
\mbf{A}_0^{[i]} = \begin{bmatrix} \mbf{A}_z^{[i]} \\ \check{\mbf{A}}_z^{[i]} \mbf{G}_{0}^{[i]} \end{bmatrix}, \; \mbf{b}_0^{[i]}(\mbf{u}_0) = \begin{bmatrix} \mbf{b}_z^{[i]} \\ -\check{\mbf{A}}_z^{[i]} \mbf{c}_0^{[i]} - \check{\mbf{B}}^{[i]} \mbf{u}_{0} \end{bmatrix}.
\end{equation}

Define the solution mappings $(\bm{\phi}_k^{[i]},\bm{\psi}_k^{[i]}) : \realset^{(k+1)n_u} \times \realset^n \times \realset^{(k+1)n_w} \times \realset^{n_v} \to \realset^{n+n_w} \times \realset^{n_y}$ and the state and output reachable tubes as in Section \ref{sec:AFDLZ}. As in \cite{Rego2020}, using the CZ version of \eqref{eq:desc_lzlimage}--\eqref{eq:desc_lzmsum}, and taking note that by assumption $\check{\mbf{z}}_k^{[i]} \in (\check{\mbf{G}}_\text{A}^{[i]}, \check{\mbf{c}}_\text{A}^{[i]}, \mbf{A}_\text{A}, \mbf{b}_\text{A})_\text{CZ}$ for every $k \geq 0$, with the latter being the lower part of $Z_\text{A}^{[i]}$, the set $Z_k^{[i]}(\seq{\mbf{u}})$ is given by the CZ $( \mbf{G}_k^{[i]}, \mbf{c}_k^{[i]}(\seq{\mbf{u}}), \mbf{A}_k^{[i]}, \mbf{b}_k^{[i]}(\seq{\mbf{u}}))_\text{CZ},$ 
where $\mbf{G}_k^{[i]}$, $\mbf{c}_k^{[i]}(\seq{\mbf{u}})$, $\mbf{A}_k^{[i]}$, and $\mbf{b}_k^{[i]}(\seq{\mbf{u}})$ are obtained by the recursive relations
\begin{equation} \label{eq:desc_recursivestatecz}
\begin{aligned}
& \mbf{c}_k^{[i]}(\seq{\mbf{u}}) = \!\begin{bmatrix} \tilde{\mbf{A}}_z^{[i]} \mbf{c}_{k-1}^{[i]}(\seq{\mbf{u}}) + \tilde{\mbf{B}}^{[i]} \mbf{u}_{k-1} \\ \check{\mbf{c}}_\text{A}^{[i]} \end{bmatrix}\!, \mbf{G}_k^{[i]} = \begin{bmatrix} \tilde{\mbf{A}}_z^{[i]} \mbf{G}_{k-1}^{[i]}\!\! & \mbf{0} \\ \mbf{0} & \check{\mbf{G}}_\text{A}^{[i]}\end{bmatrix}\!, \\
& \mbf{A}_k^{[i]} = \begin{bmatrix} \mbf{A}_{k-1}^{[i]} & \mbf{0} \\ \mbf{0} & \mbf{A}_\text{A}^{[i]} \\ \multicolumn{2}{c}{\check{\mbf{A}}_z^{[i]} \mbf{G}_{k}^{[i]}} \end{bmatrix}, \mbf{b}_k^{[i]}(\seq{\mbf{u}}) = \begin{bmatrix} \mbf{b}_{k-1}^{[i]}(\seq{\mbf{u}}) \\  \mbf{b}_\text{A}^{[i]} \\ -\check{\mbf{A}}_z^{[i]} \mbf{c}_k^{[i]}(\seq{\mbf{u}}) - \check{\mbf{B}}^{[i]} \mbf{u}_{k} \end{bmatrix},  
\end{aligned}
\end{equation}
for $k = 1,2,\ldots,N$.
The third set of constraints in $(\mbf{A}_k^{[i]}$, $\mbf{b}_k^{[i]}(\seq{\mbf{u}}))$ comes from the fact that \eqref{eq:desc_systemSVDfaultconstraintslz} must hold. Using the initial values \eqref{eq:initialAb} and following the recursive equations \eqref{eq:desc_recursivestatecz}, the state reachable tube $\seq{Z}\zerospace^{[i]}(\seq{\mbf{u}})$ can be written as an explicit function of the input sequence $\seq{\mbf{u}}$ as $\seq{Z}\zerospace^{[i]}(\seq{\mbf{u}}) = (\seq{\mbf{G}}\zerospace^{[i]}, \seq{\mbf{c}}\zerospace^{[i]}(\seq{\mbf{u}}), \seq{\mbf{A}}\zerospace^{[i]}, \seq{\mbf{b}}\zerospace^{[i]}(\seq{\mbf{u}}) )_\text{CZ}$, where $\seq{\mbf{c}}\zerospace^{[i]}(\seq{\mbf{u}}) = \seq{\mbf{Q}}\zspace^{[i]} \mbf{c}_z^{[i]} + \seq{\mbf{p}}\zspace^{[i]} + \seq{\mbf{H}}\zerospace^{[i]} \seq{\mbf{u}}$,  $\seq{\mbf{G}}\zerospace^{[i]} = [ \seq{\mbf{Q}}\zspace^{[i]} \mbf{G}_z^{[i]} \,\; \zseq{\mbf{P}}_\text{G}^{[i]}]$,  $\seq{\mbf{b}}\zerospace^{[i]}(\seq{\mbf{u}}) = \seq{\bm{\alpha}}\zerospace^{[i]} + \seq{\bm{\Lambda}}\zerospace^{[i]} \mbf{c}_z + \seq{\bm{\Omega}}\zerospace^{[i]} \seq{\mbf{u}}$, and $\zseq{\mbf{A}}^{[i]}$ given by \eqref{eq:tubelz_A}, with $\seq{\bm{\alpha}}\zerospace^{[i]} \triangleq \seq{\bm{\beta}}\zerospace^{[i]} + \seq{\bm{\Upsilon}}\zerospace^{[i]} \seq{\mbf{p}}\zspace^{[i]}$, $\zseq{\bm{\Lambda}}^{[i]} \triangleq \zseq{\bm{\Upsilon}}^{[i]} \zseq{\mbf{Q}}^{[i]}$, $\zseq{\bm{\Omega}}^{[i]} \triangleq \zseq{\bm{\Gamma}}^{[i]} + \zseq{\bm{\Upsilon}}^{[i]} \zseq{\mbf{H}}^{[i]}$, and the variables $\zseq{\mbf{p}}^{[i]}$, $\zseq{\mbf{Q}}^{[i]}$, $\zseq{\mbf{H}}^{[i]}$, $\zseq{\bm{\beta}}^{[i]}$, $\zseq{\bm{\Upsilon}}^{[i]}$, $\zseq{\bm{\Gamma}}^{[i]}$, $\zseq{\mbf{P}}_\text{G}^{[i]}$ given by \eqref{eq:tubelz_p}--\eqref{eq:tubelz_PG}, with $\mbf{G}_\text{A} \triangleq \check{\mbf{G}}_\text{A}$, and $\mbf{c}_\text{A} \triangleq \check{\mbf{c}}_\text{A}$.

The output reachable tube is obtained analogously to the LZ case, as $\zseq{Y}_\text{CZ}^{[i]}(\seq{\mbf{u}}) = \seq{\mbf{F}}\zerospace^{[i]} \seq{Z}\zerospace^{[i]} \oplus \seq{\mbf{D}}\zerospace^{[i]} \seq{\mbf{u}} \oplus \seq{\mbf{D}}\zerospace^{[i]}_v \seq{V}$. Using the CZ versions of \eqref{eq:desc_lzlimage} and \eqref{eq:desc_lzmsum}, and letting $V = ( \mbf{G}_v, \mbf{c}_v, \mbf{A}_v,$ $ \mbf{b}_v )_\text{CZ}$, this set is given by $\zseq{Y}_\text{CZ}^{[i]}(\seq{\mbf{u}}) = (\zseq{\mbf{G}}^{Y[i]}, \zseq{\mbf{c}}^{Y[i]}(\seq{\mbf{u}}), \zseq{\mbf{A}}^{Y[i]}, \zseq{\mbf{b}}^{Y[i]}(\seq{\mbf{u}}) )_\text{CZ}$, where
\begin{equation} \label{eq:outputtubecz}
\begin{aligned}
& \zseq{\mbf{c}}^{Y[i]}(\seq{\mbf{u}}) = \zseq{\mbf{F}}^{[i]} \zseq{\mbf{c}}^{[i]} + \zseq{\mbf{D}}^{[i]} \seq{\mbf{u}} + \zseq{\mbf{D}}^{[i]}_v \seq{\mbf{c}}_v, 
\\ & \seq{\mbf{G}}\zerospace^{Y[i]} = \big[\seq{\mbf{F}}\zerospace^{[i]} \seq{\mbf{G}}\zerospace^{[i]} \,\; \seq{\mbf{D}}\zerospace^{[i]}_v \seq{\mbf{G}}_v \big], \; \seq{\mbf{A}}\zerospace^{Y[i]} = \text{bdiag}(\seq{\mbf{A}}\zerospace^{[i]}, \seq{\mbf{A}}_v), \\
& \seq{\mbf{b}}\zerospace^{Y[i]}(\seq{\mbf{u}}) = [ (\seq{\mbf{b}}\zerospace^{[i]}(\seq{\mbf{u}}))^T \,\; (\seq{\mbf{b}}_v)^T]^T, \\
\end{aligned}
\end{equation}
with $\seq{\mbf{F}}\zerospace^{[i]} \triangleq 
\text{bdiag}(\mbf{F}^{[i]}, \ldots, \mbf{F}^{[i]})$, $\seq{\mbf{D}}\zerospace^{[i]} \triangleq 
\text{bdiag}(\mbf{D}^{[i]}, \ldots, \mbf{D}^{[i]})$, $\seq{\mbf{D}}\zerospace^{[i]}_v \triangleq 
\text{bdiag}(\mbf{D}^{[i]}_v, \ldots, \mbf{D}^{[i]}_v)$, $\seq{\mbf{F}}\zerospace^{[i]} = \text{bdiag}(\mbf{F}^{[i]}, \ldots, \mbf{F}^{[i]})$, $\mbf{L} = [ \eye{n} \,\; \mbf{0}_{n \times n_w}]$, $\mbf{F}^{[i]} = \mbf{C}^{[i]} \mbf{T}^{[i]} \mbf{L}$, $\seq{\mbf{c}}_v = (\mbf{c}_v, \ldots, \mbf{c}_v)$,  $\seq{\mbf{G}}_v = 
\text{bdiag}(\mbf{G}_v, \ldots, \mbf{G}_v)$,  $\seq{\mbf{A}}_v = 
\text{bdiag}(\mbf{A}_v^{[i]}, \ldots, \mbf{A}_v^{[i]})$, and $\seq{\mbf{b}}_v = (\mbf{b}_v, \ldots, \mbf{b}_v)$.

Consider the input sequence $\seq{\mbf{u}} \in \seq{U}$ to be injected into the set of models \eqref{eq:desc_systemSVDfaultlz}, and let $\seq{\mbf{y}}\zspace^{[i]} \triangleq (\mbf{y}_0^{[i]}, \mbf{y}_1^{[i]}, \ldots, \mbf{y}_N^{[i]})$ denote the observed output sequence of model $i \in \modelset$. We are interested in the design of an input sequence satisfying $\seq{\mbf{u}} \in \mathcal{S}(\zseq{Y}_\text{CZ}^{[\cdot]}(\seq{\mbf{u}}),\modelset,\seq{U})$. 

\begin{theorem} \rm \label{thm:separatinginput}
	An input $\seq{\mbf{u}} \in \seq{U}$ belongs to $\mathcal{S}(\zseq{Y}_\text{CZ}^{[\cdot]}(\seq{\mbf{u}}),\modelset,\seq{U})$ iff
\begin{equation} \label{eq:separationconditioncz}
 \begin{bmatrix} \zseq{\mbf{N}}^{(q)} \\ \zseq{\bm{\Omega}}^{(q)} \end{bmatrix} \seq{\mbf{u}} \notin \zseq{Y}^{(q)} \triangleq \left( \begin{bmatrix} \zseq{\mbf{G}}^{Y(q)} \\ \zseq{\mbf{A}}^{Y(q)} \end{bmatrix}, \; \begin{bmatrix} \zseq{\mbf{c}}^{Y(q)} \\ -\zseq{\mbf{b}}^{Y(q)} \end{bmatrix} \right)_\text{Z}
\end{equation}	
$\forall q \in \mathcal{Q}$, where $\zseq{\mbf{N}}^{(q)} \triangleq (\zseq{\mbf{F}}^{[j]} \zseq{\mbf{H}}^{[j]} + \zseq{\mbf{D}}^{[j]}) - (\zseq{\mbf{F}}^{[i]} \zseq{\mbf{H}}^{[i]} + \zseq{\mbf{D}}^{[i]})$, $\zseq{\bm{\Omega}}^{(q)} \triangleq [(\zseq{\bm{\Omega}}^{[i]})^T \; \mbf{0} \; (\zseq{\bm{\Omega}}^{[j]})^T \; \mbf{0}]^T$, $\zseq{\mbf{G}}^{Y(q)} \triangleq [\zseq{\mbf{G}}^{Y[i]} \,\; - \zseq{\mbf{G}}^{Y[j]}]$, $\zseq{\mbf{c}}^{Y(q)} \triangleq \zseq{\mbf{c}}^{Y[i]} (\mbf{0}) - \zseq{\mbf{c}}^{Y[j]} (\mbf{0})$, $\zseq{\mbf{A}}^{(q)} \triangleq \text{bdiag}(\zseq{\mbf{A}}^{Y[i]},\zseq{\mbf{A}}^{Y[j]})$, and $\zseq{\mbf{b}}^{Y(q)} \triangleq [(\zseq{\mbf{b}}^{Y[i]}(\mbf{0}))^T \,\; (\zseq{\mbf{b}}^{Y[j]}(\mbf{0}))^T]^T$, in which  $i,j \in \modelset$ is the pair associated to each $q \in \mathcal{Q}$, and $\seq{\mbf{0}}$ denotes the zero input sequence.
\proof The proof is similar to Theorem \ref{thm:desc_separatinginputlz}, and therefore is omitted. \qed
\end{theorem}

Since $\zseq{Y}^{(q)}$ given by \eqref{eq:separationconditioncz} is a zonotope, analogously to the methods in \cite{Scott2014} and \cite{Rego2020b}, the relation \eqref{eq:separationconditioncz} can be verified by solving the LP
\begin{equation} \label{eq:separationconditionczLP}
\hat{\delta}^{(q)}(\seq{\mbf{u}}) \triangleq \left\{ \underset{\delta^{(q)}, \bm{\xi}^{(q)}}{\min} \delta^{(q)} : \left. \begin{aligned} & \begin{bmatrix} \zseq{\mbf{N}}^{(q)} \\ \zseq{\bm{\Omega}}^{(q)} \end{bmatrix} \seq{\mbf{u}} = \begin{bmatrix} \zseq{\mbf{G}}^{Y(q)} \\ \zseq{\mbf{A}}^{Y(q)} \end{bmatrix} \bm{\xi}^{(q)} + \begin{bmatrix} \zseq{\mbf{c}}^{Y(q)} \\ -\zseq{\mbf{b}}^{Y(q)} \end{bmatrix}, \\ & \ninf{\bm{\xi}^{(q)}} \leq 1 + \delta^{(q)} \end{aligned} \right. \right\},
\end{equation}
where \eqref{eq:separationconditioncz} holds iff $\hat{\delta}^{(q)}(\seq{\mbf{u}}) > 0$. 

\begin{remark} \rm
    Similarly to the LZ case, the complexity of the LP \eqref{eq:separationconditionczLP} can be reduced by using the change of variables $\zseq{\mbf{u}} = \zseq{\mbf{c}} + \zseq{\mbf{G}} \zseq{\bm{\xi}}_u$, and applying the constraint elimination approach described in Section \ref{sec:inputconstraintelimlz} to CZs.
\end{remark}

Finally, consider the design of an input sequence that minimizes a functional $J(\seq{\mbf{u}})$ subject to $\seq{\mbf{u}} \in \mathcal{S}(\seq{Y}\zspace_\text{CZ}^{[\cdot]}(\seq{\mbf{u}}),\modelset,\seq{U})$, with initial conditions $X_0$ and $\mbf{u}_0$. As in the LZ method, if $J(\seq{\mbf{u}}) \triangleq (\seq{\mbf{u}} - \tilde{\mbf{u}})\zspace^T \mbf{R} (\seq{\mbf{u}} - \tilde{\mbf{u}})$, where $\mbf{R} \in \realsetmat{(N+1)n_u}{(N+1)n_u}$ is positive definite, and $\tilde{\mbf{u}} \in \seq{U}$ is a reference input sequence, then this optimization problem, similar to \eqref{eq:fault_optimalseparatinglzfinal}, can be written as an MIQP. On the other hand, if $J(\seq{\mbf{u}}) = \|\mbf{R} (\seq{\mbf{u}}-\tilde{\mbf{u}})\|_1$, it leads to an MILP.

\bibliography{\bibfolder/masterthesis_bib,\bibfolder/appendices_bib,\bibfolder/UAVControl_bib,\bibfolder/BackgroundHist_bib,\bibfolder/Surveys_bib,\bibfolder/PassiveFTC_bib,\bibfolder/ActiveFTC_bib,\bibfolder/UAVFTC,\bibfolder/SetTheoretic_bib,\bibfolder/SetTheoreticFTCFDI_bib,\bibfolder/Davide_bib,\bibfolder/paperAutomatica_bib,\bibfolder/paperCDC_bib,\bibfolder/paperECC_bib,\bibfolder/paperIFAC_bib,\bibfolder/paperNonlinearMeas_bib,\bibfolder/Robotic_bib,\bibfolder/stelios_bibliography,\bibfolder/phdthesis_bib,\bibfolder/paperParameter_bib,\bibfolder/Diego_bib,\bibfolder/paperMixed_bib}

\end{document}